\documentclass{ws-ijmpa}
\def\beq{\begin{equation}}

\def\eeq{\end{equation}}

\begin{document}

\markboth{A. Kovner, M. Lublinsky}
{Angular and long range rapidity correlations in particle production at high energy.}

%\catchline{}{}{}{}{}
%
%%%%%%%%%%%%%%%%%%%%%%%%%%%%%%%%%%%%%%%%%%%%%%%%%%%%%%%%%%%%%%%%%%%%

\title{ANGULAR AND LONG RANGE RAPIDITY CORRELATIONS IN PARTICLE PRODUCTION AT HIGH ENERGY}

\author{ALEX KOVNER}

\address{Physics Department, University of Connecticut, 2152 Hillside
Road, \\ Storrs, CT 06269-3046, USA\\ kovner@phys.uconn.edu}

\author{MICHAEL LUBLINSKY}

\address{Physics Department, Ben-Gurion University of the Negev, \\ Beer Sheva 84105, Israel\\ lublinm@bgu.ac.il}

\maketitle

%\begin{history}
%\received{Day Month Year}
%\revised{Day Month Year}
%\end{history}

\begin{abstract}
We discuss the general mechanism leading to long-range rapidity and angular correlations produced in high energy collisions (the "ridge"). This effect naturally appears  in the high energy QCD and is strongly sensitive to  physics of the gluon saturation. We comment on various recent practical realizations of the main idea, paying
special attention  to  $N_c$ counting and  stress the relevance of Pomeron loops.
\end{abstract}

%\ccode{PACS numbers: ???????}

\section{Introduction}

While built as a discovery machine,  the LHC at CERN  has a very rich research program  exploring QCD in extreme conditions.
This program ranges from the fundamental measurement of total scattering cross section in proton-proton (pp) collisions at very high energies
to production and study of Quark Gluon Plasma (QGP) in heavy ion collisions.  

One of the first and intriguing observations made by the CMS collaboration\cite{CMS} is a discovery of  long-range  two-particle 
correlations in $pp$ collisions at $\sqrt s= 7\,TeV$.  The correlations are observed between two charged hadrons which emerge from the collision
collimated in their azimuthal angle $\phi$ while being widely separated in (pseudo-) rapidity $\eta$.  Remarkably, 
while not present in the minimal bias events, the correlations are observed in high multiplicity events. These are extremely rare events
(approximately $10^{-6}$ of the total event sample) in which the total multiplicity of produced charged hadrons is many (7-8) times larger than a typical multiplicity
in minimal bias collisions.

Correlations, similar to the ones observed by the CMS, were seen before at RHIC in collisions of heavy ions\cite{STAR,PHOBOS,star1}. 
The phenomenon was dubbed "ridge"
due to its shape on the $\phi-\eta$ data plot (see below).  An accepted explanation of the "ridge" at RHIC is related to collective flow in QGP. 
The fact that a similar ridge structure has been observed in $pp$ collisions may hint at an intriguing possibility that high multiplicity $pp$ events produce a state somewhat similar to QGP.  This question is however still open today, as much more data would be needed to provide a clear answer.

Recently, CMS released its preliminary data from the test run of proton-lead collisions at $5.02\,TeV$ per nucleon\cite{pA}. The "ridge"-like structure
of di-hadron azimuthal correlations is again clearly seen in high multiplicity events, and its magnitude is significantly stronger than in the $pp$ data.

There is a theoretical basis for expectation that at asymptotically high energies hadrons and nuclei behave alike.
At high energies both protons and heavy ions appear as dense clouds of gluons. The growth of  gluon densities with energy is governed by the QCD evolution equations, which are universal for all hadrons.  Thus, from this point of view, 
$pp$ collisions at very high energies  are expected  to resemble heavy ion collisions  at somewhat smaller energy. It is however overly optimistic to expect that this regime has already been reached at LHC, and it certainly has not been reached for the minimum bias $pp$ collisions.  The rare high multiplicity events may represent the tail of the distribution where high density has been achieved already, but this too is far from certain. 

The long range rapidity correlations may not be (and probably are not) related to the formation of QGP. However they are a striking feature of the observed data, and are very likely a universal property of a highly energetic QCD collision, provided the number of particles produced in the final state is large enough to probe such correlations. Since the correlations extend over a very large rapidity interval, causality considerations\cite{DMV} (similar to cosmological arguments behind
baryonic acoustics and CMB fluctuations) lead one to conclusion that the origin of the correlations is at very early stage of the collision, and in fact most likely in the initial state. Without such early time correlations, relativistic particles separated by many units of rapidity flying out of the interaction region would  be causally disconnected and uncorrelated.   
At the same time, high multiplicities of produced particles must be due to
high gluon densities of colliding hadrons. We are thus  naturally lead to a conclusion that the observed correlations are rooted in the 
high density QCD (hdQCD). The nature of this relation is the subject of this review paper.

There have been numerous publications proposing various mechanisms responsible for the "ridge" in AA and  in $pp$ collisions\cite{recent1,recent2,recent3,recent4,recent5,recent6,recent7}. 
It is not our goal  to provide a full coverage of those. Here  we focus on correlations that originate in the initial conditions, or the  wave-functions of colliding hadrons and their relation to high density effects.  The main ideas reviewed here, appeared first in Refs.(\refcite{ddgjlv,correlations,LR}).  We will explain how the theory of high density QCD  naturally leads to the correlations of the kind observed in the $pp$ experimental data. The mechanism discussed here does not refer at all to collective flow in the final state, and thus is not likely to be relevant for the explanation of the ridge in heavy ion collisions. We take in this review the point of view that although the origin of long range rapidity correlations is similar in heavy ion and proton-proton collisions, the angular collimation in the two systems is probably of a different origin. 

hdQCD manifests itself in the phenomenon of gluon density saturation \cite{GLR}. Gluon densities grow with energy. The low (Bjorken) $x$ evolution of gluon densities 
is governed by the linear BFKL equation\cite{BFKL,BFKL1}, which leads to a rapid (exponential with rapidity) growth.
Once the densities reach the "critical" value, where the density of gluons per unit phase space is of order of the inverse of the strong coupling constant, non-linear effects in the evolution become important. 
These effects slow down the growth of the density, leading to gluon saturation, characterized 
by the appearance  of a new scale $Q_s(x)$, known as the saturation momentum.  
Such a dense gluonic  state is frequently referred to as  the Color Glass Condensate (CGC)\cite{cgc,cgc1,cgc2}.
The strong coupling constant $\alpha_s$,  evaluated at  the saturation scale 
$Q_s^2\simeq 5\,GeV^2$ for the LHC energies, is rather small, which makes many aspects of saturation perturbatively accessible.  However, even for a small coupling constant, the fact that the gluon densities are high invalidates naive perturbation theory. The proper treatment of the saturation regime therefore requires resummation of the naive perturbation theory. 

The nonlinear generalization  of  the BFKL equation that 
resums the high 
density effects is
known as the BK\cite{balitsky,balitsky1,balitsky2,Kovchegov}-
JIMWLK\cite{JIMWLK,JIMWLK1,JIMWLK2}\cite{JIMWLK3,JIMWLK4,JIMWLK5,cgc,cgc1,cgc2}
 functional equation.
 This (functional) equation is strictly speaking applicable when one of the colliding particles is dilute,
 such as in the deep inelastic scattering (DIS) explored at  HERA, or in proton-nucleus collisions.
Although this is not the case when both colliding particles are high energy hadrons, and the QCD description of such systems is still not well understood, we will be using the BK-JIMWLK approach as a proxy to qualitative description of hadronic high energy scattering. 

%The advantage of two- and multi-particle correlation studies is that these observables provide a much richer insight about mechanisms of 
%particle production beyond what one could infer  from single inclusive measurements.

Some aspects of relation of saturation physics with the ridge phenomenon have been reviewed in Refs.(\refcite{Li,Kovner,Lappi,Iancu,Raju}).

\section{Experimental data on di-hadron correlations}

\subsection{The "Ridge" in heavy ion collisions.}

\begin{figure}[pb]
\centerline{\psfig{file=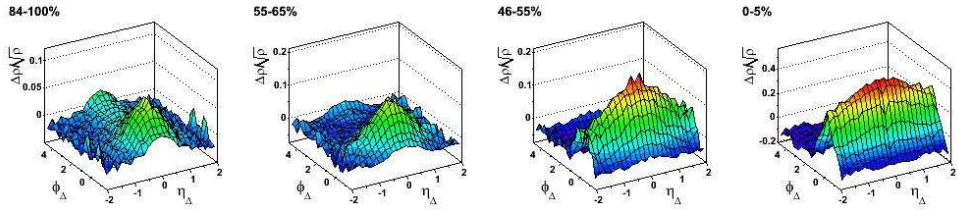,width=13cm}}
\vspace*{8pt}
\caption{Soft ridge from STAR
%\cite{STAR}. 
The centrality dependence. \label{f1}}
\end{figure}

\begin{figure}[pb]
\centerline{\psfig{file=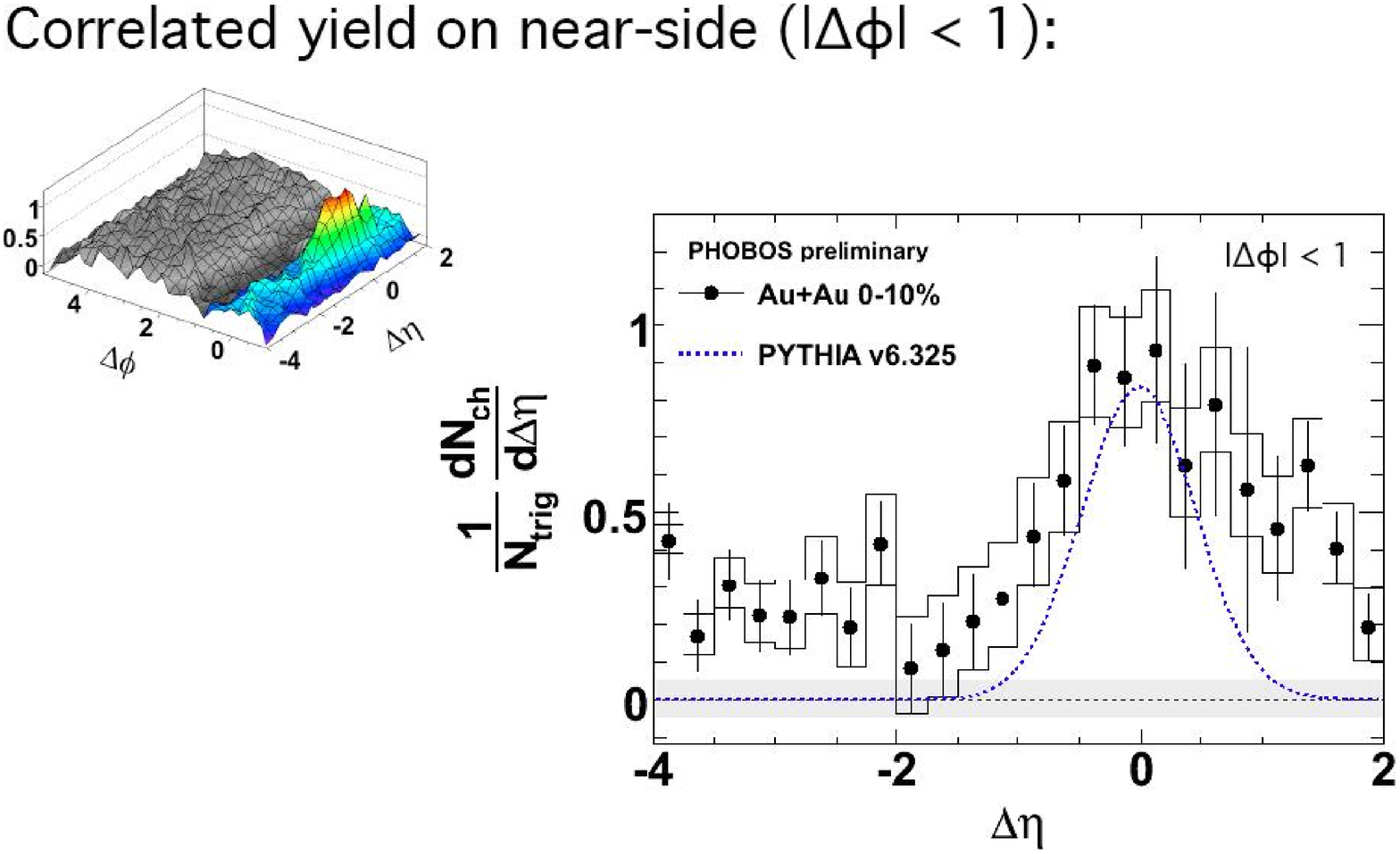,width=13cm}}
\vspace*{8pt}
\caption{Hard ridge from PHOBOS
%\cite{PHOBOS}. 
\label{f2}}
\end{figure}

%The main new feature that is characteristic to heavy ion collisions and not present in any other collider experiments is emergence of
%collective flows. When talking about two-particle correlations, the collective flows, such as elliptic flow $v_2$, are normally subtracted
%from the data, leaving the net correlations exposed. 
The ridge was first observed in gold-gold collisions at $\sqrt s=200\, GeV$ 
at RHIC by the STAR experiment\cite{STAR}  (Fig.~\ref{f1}). The initial measurement of di-hadron correlations was performed with a high momentum trigger particle $p_T> 4\ GeV$, but subsequently the correlations were measured for both, trigger and associated momenta in the (semi) soft range.
Here $\phi$ is the azimuthal angle, $\Delta\phi$  is the angular difference between the momentum of the measured hadron
and the direction of the trigger.  The peak around $\Delta\phi=0$ corresponds to the trigger jet fragmentation region.  The bump at $\Delta\phi=\pi$
is the usual back-to-back correlation. $\eta\simeq ln(p_z/p_\perp)$ stands for (pseudo-)rapidity of produced particles.  For two hadrons with approximately 
equal transverse momenta, $\Delta\eta$ measures their relative momenta along the beam direction, or, in other words,  relative energies.
Centrality dependence (indicated in percents above each plot) of the "ridge"  clearly demonstrates the ridge in central collisions while it disappears
towards peripheral region.  Further studies by PHOBOS\cite{PHOBOS}  and STAR \cite{star1}(Fig.~\ref{f2}) were performed with and without hard trigger and confirmed the appearance of the ridge and its essential (semi) soft nature.

\begin{figure}[pb]
\vspace*{-4cm}
\centerline{\psfig{file=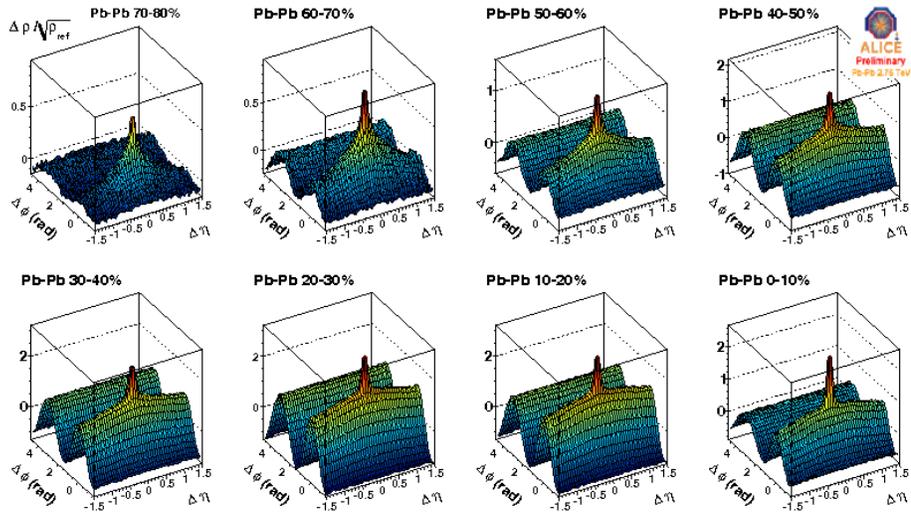,width=13cm}}
\vspace*{-4.5cm}
\caption{Soft ridge from ALICE
%\cite{ALICE}. 
\label{Alice}}
\end{figure}

Lead-Lead collisions at $\sqrt s=2.76\, TeV$ at the LHC provide further information about the ridge in heavy ion collisions. With  increase in collision
energy, the $p_t$ range of particles that exhibit ridge-type correlations broadens significantly. CMS has observed  the ridge up to $p_t\simeq \,20\, GeV$\cite{CMSnew}. 
ALICE has also studied the soft ridge\cite{ALICE,ALICE1} and their results on centrality dependence are shown on Fig.~\ref{Alice}.

The standard explanation of the ridge in HI collisions refers to collective flow\cite{shuryak,voloshin}. 
Particularly, angular collimation in $\phi$ can be attributed to the radial flow that boosts particles along the radial direction\cite{GMM,DL3}.  
%More refined analyses of the data, based on Fourier decomposition with respect to $\Delta\phi$, include radial and elliptic flow
%components as well as higher (up to six) Fourier harmonics.  The success of phenomenological fits to reproduce the ridge data
%can be taken as a convincing argument in favor of strong fluctuations at early stage of fireball evolution.

\subsection{Angular correlations in $pp$ @CMS}

One of the first pieces of data to come from LHC was the observation of ridge in $pp$ collisions by the CMS experiment.
\begin{figure}[pb]
\centerline{\psfig{file=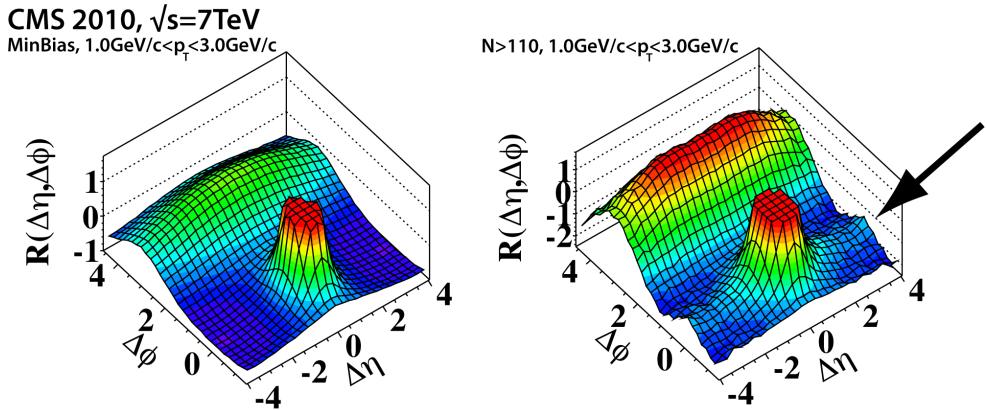,width=13cm}}
\vspace*{8pt}
\caption{Di-hadron correlations from CMS
%\refcite{CMS}. 
Left: minimal bias data; Right: high multiplicity data.   \label{f3}}
\end{figure}

Fig.~(\ref{f3}) displays  the CMS data at $7\,TeV$.  
The left plot displays a typical minimal bias event. The right plot shows data on specially selected high multiplicity events.  The pronounced "ridge"-like
structure in the near side ($\Delta\phi\simeq 0$) region extending well beyond the jet fragmentation rapidity  interval all the way to $\Delta\eta=4$  is a distinct 
correlated feature.  It has never been observed in $pp$ before, nor can it be explained by any of the existing 
Monte Carlo generators used to simulate hadron cascades.

As was mentioned in the Introduction, only one out of approximately $10^{6}$ events leads to extraordinary high multiplicities
produced in the final state. It is this set of events that exhibits the ridge like correlations. Importantly, the correlations are observed only between semi-hard particles with transverse momenta limited to the interval $1GeV<p_t< 3GeV$. High density QCD naturally singles out this interval as being related to gluon saturation scale $Q_s\simeq 2GeV$, typical for such events.

Following the initial  measurement\cite{CMS}, further CMS studies \cite{CMSnew} have revealed  that the ridge gets stronger with increase of 
final multiplicities: the ridge emerges at multiplicities of order $N\sim 50$ and becomes more pronouned at 7-8 times the average multiplicity. 
%This observation provides further support for the statement made in the Introduction that the phenomenon is a high density effect.
 
%The width of ridge is an interesting observable. It provides information about origin of collimation. What else???
%?Was it studies?? To what it is sensitive to? 
Although multiplicities in the relevant events are large, they are still very far off the typical multiplicities in $Au-Au$ collisions at RHIC at $200\ GeV$. It is unlikely that radial flow plays a significant role in these events, and we will take in this review the point of view that final state interactions do not change qualitatively correlations produced during the collision. If this is the case, the forward angular correlations observed by CMS require an independent explanation.

\subsection{Ridge in $pPb$ @CMS}

\begin{figure}[pb]
\psfig{file=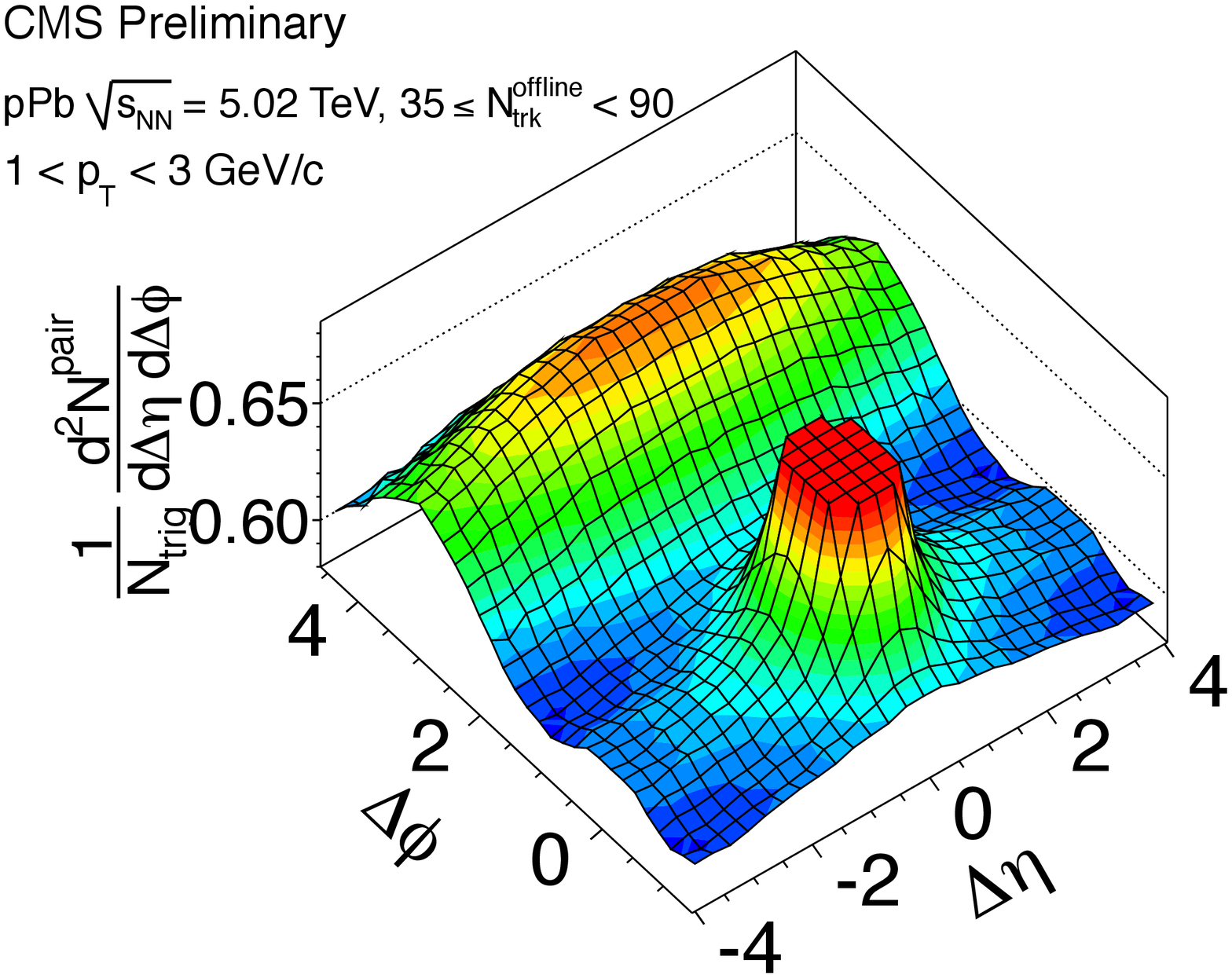,width=4.5cm}\hspace*{-0.3cm}
\psfig{file=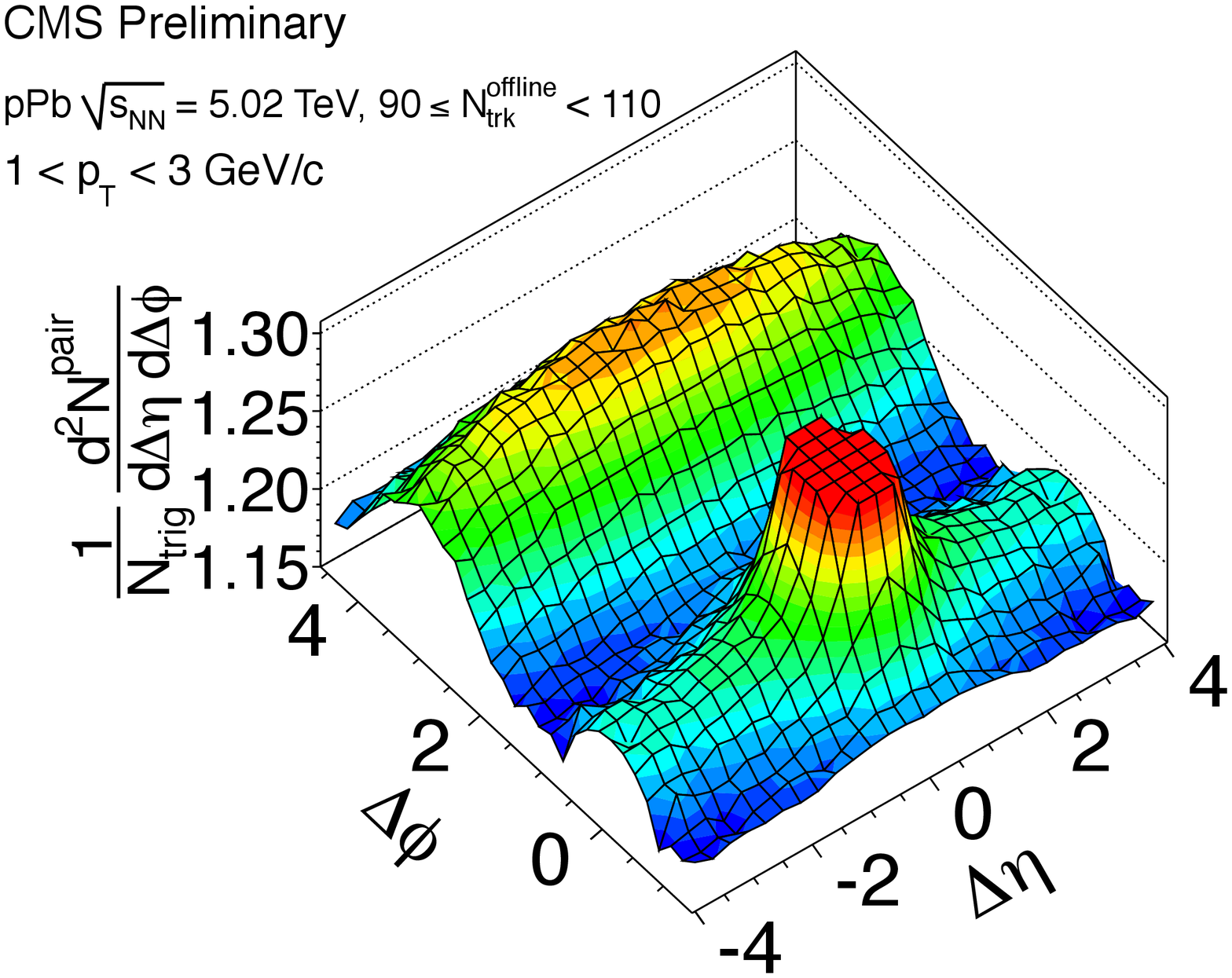,width=4.5cm}\hspace*{-0.3cm}
\psfig{file=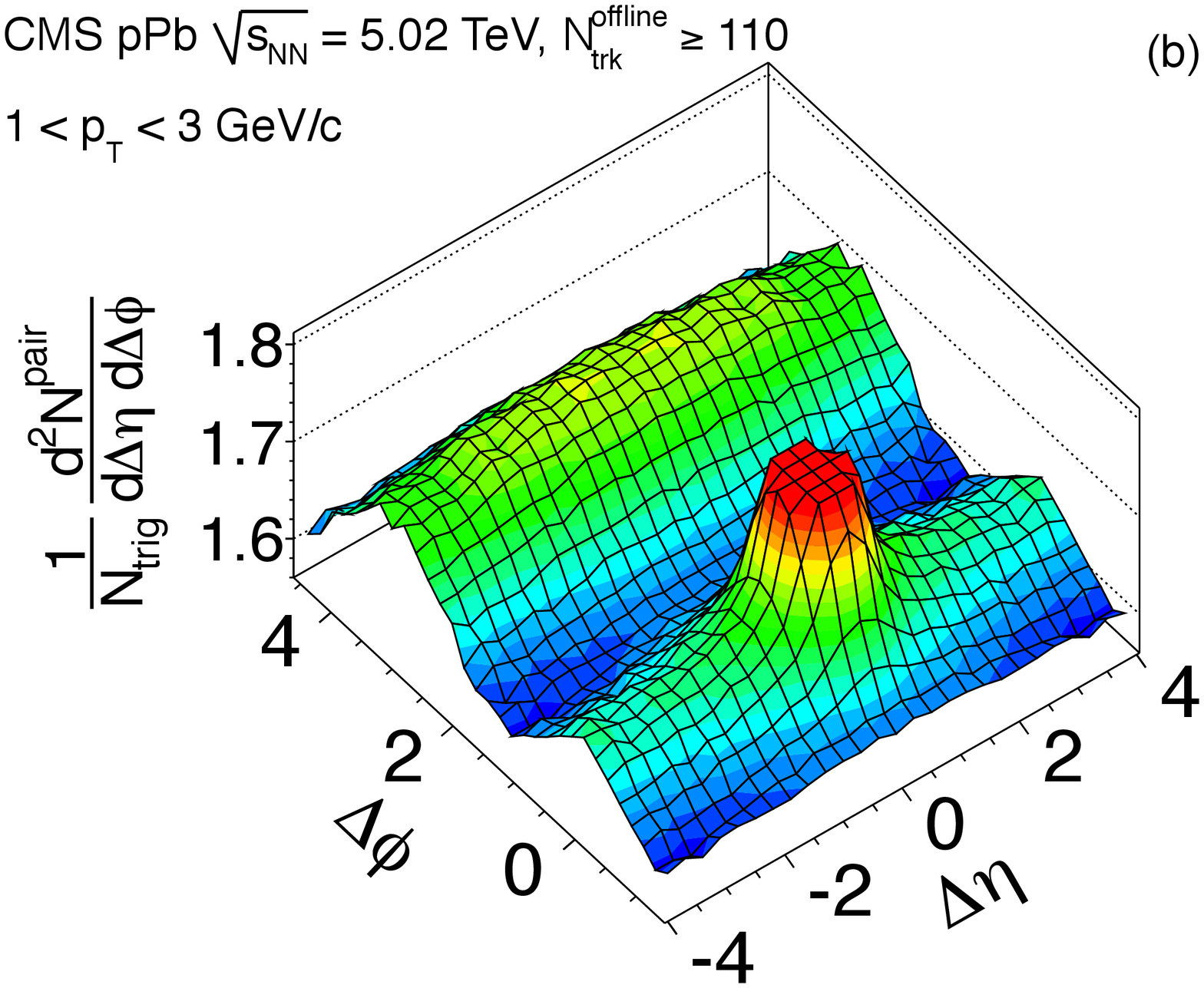,width=4.5cm}
\vspace*{-15mm}
\caption{
%\cite{STAR}. 
Multiplicity dependence of the ridge in $pPb$. \label{pPb}}
\end{figure}

Preliminary CMS data on the proton-lead collisions at $5.02\,TeV$ from the LHC test run  have been released recently\cite{pA}.
Analysis similar to one performed by the CMS in $pp$ case again revealed the ridge-like structure in the semi-soft momenta interval $1<p_t<3\,GeV$.
Special focus given to the ridge's strength (height) as a function of the total multiplicity of produced charged hadrons, reveals an approximately linear
relation (Fig. \ref{pPb}), while the overall strength of the ridge is found to be mach more pronounced than in the $pp$ case. 

\section{Correlations from Theory}

\subsection{General considerations}

To explain the origin of the experimentally observed  di-hadron correlations one has to address two distinct questions:
(a) why do the correlations extend over such a long rapidity range; and (b) what is the mechanism responsible for correlating the observed hadrons within a relatively small angle. 

\begin{figure}[pb]
\vspace*{-3cm}
\centerline{\psfig{file=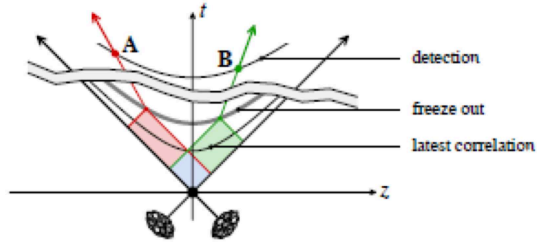,width=8cm}}
\vspace*{-2.8cm}
\caption{Causality structure of correlations
%\cite{DMV}. 
\label{caus}}
\end{figure}

As has been argued in Ref. (\refcite{DMV}), causality strongly constrains the time at which the correlations originate. In order for two hadrons separated by a large rapidity interval to be
correlated at the time of detection, they must be correlated at very early times: at or just after the collision. As can be seen from the Fig. ~(\ref{caus}), in  case of HI
collisions, the maximal correlation proper time  is
\beq
\tau_{latest\, correlation}\le \tau_{freezout} \,exp(-\Delta\eta/2)
\eeq
It follows immediately that the larger the rapidity interval between the two particles, the earlier the correlation must be formed.  Furthermore, the
flatness of the ridge in $\Delta\eta$ indicates that all the correlations are produced independently of the rapidities of the particles. If this was not the case, the ridge signal would be weakening for larger rapidity separation between the
produced particles, the effect that so  far has not been observed experimentally. 

In the case of $pp$ collision, the  freezout time should be replaced by a (typically very short) time scale of 
parton hadronization. These arguments lead to the conclusion that the long range correlations are present just after the interaction 
takes place, while their production mechanism is practically rapidity independent.

In QCD there exists a very simple and natural mechanism which can produce such long range rapidity correlations - the approximate boost invariance of the wave function of a very energetic hadron. As we know from variety of experiments as well as from the BFKL theory, hadronic wave function at high energy is gluon dominated. Since gluons are vector particles, their Lorentz transformation properties are such that in a boost invariant state they are homogeneously distributed in rapidity.
Thus, the observed rapidity  independence of the correlations could be  merely a reflection of
the initial conditions  of  high energy scattering processes  as encoded in QCD.

The angular correlations between produced particles are also to be expected on very general grounds at high energy. We will now explain the qualitative picture underlying these correlations. This is very general and  does not rely on specific models of high energy evolution and/or  hadronic wave function.
In the next subsection we will provide more calculational details substantiating this qualitative discussion.

Consider high energy scattering of a hadronic projectile on a stationary target in the lab frame.  Two projectile gluons with rapidities $Y_1$ and $Y_2$ scatter on the target and are observed in the final state.  As mentioned above, at high enough energy the wave function of the projectile is approximately boost invariant. Thus the projectile gluon distribution at rapidities $Y_1$ and $Y_2$ is the same. The two gluons scatter exactly on the same target, and thus whatever happens at $Y_1$ also happens at $Y_2$. If for a particular target field configuration a gluon is likely to be produced at $Y_1$ at a certain impact parameter, a gluon is also likely to be produced at $Y_2$ at the same impact parameter.  

This is especially true in the context of the projectile wave function dominated by the large "classical" Weizsacker-Williams field, since in this case fluctuations in the wave function are small and the gluon density {\it configuration by configuration} is almost the same at all rapidities.
The hadronic wave function at high energy can be computed in QCD using light-cone Hamiltonian perturbation theory\cite{KLwmij,bubbles}:
\begin{equation}
|P \rangle =\Omega|\psi\rangle\,=\,\exp\Big\{i\int d^2xb^a_i(x)\int d\eta \left(a^{\dagger a}_i(x,\eta)+a^a_i(x,\eta)\right)\Big\}B(a,a^\dagger)|\psi\rangle \,.
\label{wf}
\end{equation}
Here $\psi$ is the wave function of valence charges, that determines the distribution of fast partons responsible for the color charge density $\rho$; $B$ is a Bogolyubov -type operator of the soft gluon fields $a$, and the Weizsacker-Williams (WW) field $b$ is given in terms of $\rho$ via classical Yang-Mills equations of motion. For small (dilute) projectile
\beq b_i^a(x)=\frac{g}{\,\pi}\,\int_y f_i(x-y)\,\rho^a(y)\,, \ \ \ \ \ \ \ \ \ \ \ \ \ \ \ \  f_i(x-y)=\frac{(x-y)_i}{(x-y)^2}\,, \ \ \ \  \ \ \ \ \ \ \ \ \ B=1.\eeq
 For large projectile the WW field is parametrically large $b\sim \frac{1}{g}$, while the Bogolyubov operator $B$ produces the fluctuations of the gluon field of order unity. 
 Thus for fixed $\rho(x)$ the gluon density fluctuates very weakly around large average value determined by the classical field
\begin{equation}
n=\langle a^\dagger a\rangle \propto b^2\sim O(\frac{1}{\alpha_s}); \ \ \ \ \ \ \ \ \langle n^2\rangle -\langle n\rangle ^2\sim 1
\end{equation}
The smallness of the fluctuations is clearly helpful. Although {\it the wave function} at different rapidities in a boost invariant projectile must be the same, the magnitude of the {\it color field} (and therefore the number of gluons) may differ at different values of $Y$ for the same configuration of the valence color charge density $\rho$, if the fluctuations in this wave function are significant. Thus in the same scattering event there may be significant differences between particle production at different rapidities. 
Still, although the quasiclassical nature of the state eq.(\ref{wf}) ensures long range rapidity correlations at large values of $\rho$, it is not absolutely necessary. Even in the presence of considerable fluctuations in the soft gluon wave function, one nevertheless would expect  positive correlations in rapidity. The only really necessary condition is that the density of incoming partons is large enough, so that there is large probability to produce more than one particle at a given impact parameter (we will quantify what we mean by "given impact parameter" shortly).
 
Thus the long range rapidity correlations come practically for free whenever the energy is high enough so that the wave function of the incoming hadron is approximately boost invariant, and there is very little in the actual dynamics of the collision that can affect this feature.  

By almost exactly the same logic we can argue that near-side angular correlations must be present between the final state gluons. Indeed, if the two gluons hit the target at the same impact parameter, their scattering amplitude is determined by the same configuration of the target field. Thus if the  first gluon is likely to be scattered with momentum $q$, the same is true for the second gluon. One therefore expects clear forward correlations for gluons that scatter at the same impact parameter. 
Of course, the two gluons will not scatter always with exactly the same momentum transfer even if they hit at exactly the same impact parameter, since even a fixed configuration of target fields corresponds to a nontrivial probability distribution of momentum transfer. Nevertheless, given that this distribution has a maximum at some particular momentum transfer, the angular correlations must be very generic.

\begin{figure}[pb]
\centerline{\psfig{file=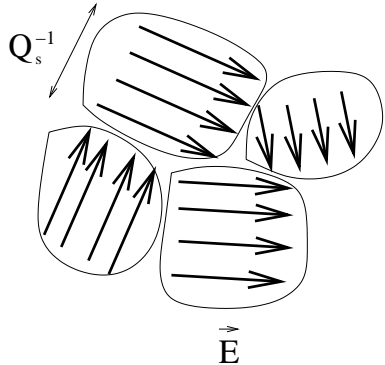,width=5cm}}
\caption{Domain structure of chromo-electric fields in the target. \label{domain}}
\end{figure}

The above argument does not invoke the physics of saturation. Nevertheless, saturation, if present in the wave function does affect strongly the gluon production.
To better understand the expected properties of the angular correlations in the framework of high energy QCD, let us briefly recap our understanding of the transverse structure of the hadron in the saturation regime. It is convenient to think of the distribution of the (color) electric field configurations in the target. 
The target wave function is characterized by the saturation momentum $Q_s$. The saturation momentum plays a dual role in the hadronic wave function. On the one hand, it measures the typical magnitude of electric field in the wave function, on which gluons scatter.  %Moreover, one should think of the saturation scale as of a {\it vector}.  
On the other hand it is known that the field components with transverse momenta $p_T<Q_s$ are suppressed in the wave function\cite{satdist,satdist1}. 
This means that the electric fields in the target are correlated on the length scale $\lambda\sim Q_s^{-1}$. Thus the saturation momentum doubles up as the inverse of the correlation length of target color fields. Typical field configurations in the target can thus be thought of having, in the impact parameter plane,
a domain like structure of  Fig. ~(\ref{domain}).
Now consider a projectile parton  impinging on one of the domains or "hot spot" of the target, with electric field  $\vec E$.
While traversing the target field, the parton of charge $q$ acquires transverse momentum
\begin{equation}
\delta \vec P= gq\int dx^+ \vec F^{-} = gq\vec E\ ; \, \ \ \ \ \ |\delta \vec P| \simeq\,  Q_s
\end{equation}
A parton at a different rapidity but with the same charge will pick up exactly the same transverse momentum if it scatters on the same "domain". 
This of course results in positive angular correlation of produced gluons.  

We note that this simple picture also suggests that angular correlations at angle $\phi$ and $\phi+\pi$ have equal strength if the hadronic wave function is dominated by gluons.
Gluons belong to real representation of the gauge group, thus it is equally probable to find an incoming gluon with charge $q$ and charge $-q$ in the projectile wave function at any rapidity. Suppose for example that on a given configuration the color field in the target is in the third direction in the color space $E^a_i=E_i\delta^{a3}$, while in the incoming projectile the gluon corresponds to the vector potential in the second direction $A^2_i$. One can always write $A^2=-i/2(A-A^*)$, where $A=A_1+iA_2$ is positively charge with respect to color charge in the third direction, and $A^-$ is negatively charged. Thus necessarily equal number of gluons in the incoming projectile have opposite sign charges and are kicked in opposite directions while scattering on the target. This produces equal strength correlations at angles zero and $\pi$. 
This feature of equal forward and backward correlations was noted in Refs.(\refcite{ddgjlv,correlations}).
For quarks which carry fundamental charges, this degeneracy should be absent and the presence of quarks in the projectile wave function should skew the angular correlation towards zero angle and away from $\Delta \phi=\pi$.

\subsection{A little more hands on.}

%In many publications  (see e.g  Refs. \refcite{GMM,DL3})  the boost invariance is addressed during a short time interval, right after the interaction instant. 
%In HI collisions, this interval extends up to the time when the system  thermalizes,  
%while the non-equilibrated state created  just after collision is frequently referred to 
%as Glasma \cite{GLASMA,GLASMA1}.
% One of the  prime features of Glasma is the presence of strong longitudinal chromo-electric fields, which are absent before the
%collision. These fields fluctuate in the transverse plane with a typical  scale of  order $Q_s$, and as such are frequently envisioned as (Glasma) fluxed tubes
%or strings stretched along the longitudinal direction (Fig.~\ref{glasma}).   
%The longitudinal fields create gluons with a probability that is independent of gluon's rapidity. 
%% \begin{figure}[pb]
 %\vspace*{-2.cm}
%\centerline{\psfig{file=tubes.ps,width=5cm}}
%\vspace*{-1.5cm}
%\caption{Glasma flux tubes
%\cite{GLASMA,GLASMA1}. 
%\label{glasma}}
%\end{figure}
% The flux tube/string picture  largely follows\cite{Alikeikonal} from  the boost invariance of  hadronic wave-functions and from the eikonal scattering approximation 
%valid at high energies. Thus, many seemingly different approaches to the problem have the same rooting into rigorous hdQCD.

The above general discussion provides an intuitive picture for the existence of correlations. Our next goal is to see how this is reflected in the calculational framework based on hdQCD. As explained above, we  will base our analysis on expressions derived rigorously at high energy  for a scattering of a dilute projectile
on a dense target. We do this for two reasons. One reason is the lack of explicit analytical expressions for the case of dense-dense scattering, although a significant advance in understanding of symmetric collisions has been achieved\cite{Kovpom,raju1,raju2,raju3,KLploop,KLploop1,Balincl} and an elegant numerical approach is being pursued in a series of papers Refs.(\refcite{raju1,raju2,raju3}). Also, as we will argue later, the dense-dense scattering approach is not entirely adequate for study of correlations either, as it does not take into account the Pomeron loop effects which in our view is crucial for quantitative understanding of correlations. The dense-dilute approach is sufficient for the understanding of basic ingredients involved in the calculation of correlations.

The quantity directly related to the measured correlations, is the double inclusive gluon production amplitude. The correlations are computed
as  variance with respect to the single inclusive measurement:
\beq\label{r}
R=\left[\frac{d^2N}{d^2p\,d{\eta}\,d^2k\,d\xi }\,-\,\frac{dN}{d^2k\,d\xi}\,\frac{dN}{ d^2p\,d\eta}\right]/\frac{dN}{ d^2k\,d\xi}\,\frac{dN}{d^2p\,d\eta}
\eeq
Here $p,\,k$ are the transverse momenta of the observed gluons, and $\eta,\,\xi$ are their respective rapidities. The experimentally measured correlations
are normalized per number of produced charged hadrons, $R^{exp}\sim R\,dN/d\xi$, however the quantity in eq.(\ref{r}) has a more straightforward interpretation.

The expressions for single- and double-inclusive gluon production probabilities are known in the literature. We outline the derivation in Appendix A, while here we quote the results.
The single-inclusive amplitude for production of a gluon with transverse momentum $k$ and rapidity $\eta$ 
reads\cite{KT,Braunincl,klw,BSV,LP}:
\begin{eqnarray}\label{single}
\frac{dN}{d^2kd\eta }= \langle \hat \sigma(k,\eta)\rangle_{P,T}&& \nonumber \\
\hat\sigma(k,\eta)&=& \frac{\alpha_s}{\pi}\int_{z,\bar z, x_{1},\bar x_{1}}e^{ik(z-\bar z)}\vec{f}( \bar z - \bar x_1)\cdot \vec{f}( x_1- z)\,\times\nonumber \\
&\times&\left\{ \rho_p^a(x_1)[S^\dagger( x_1)-S^\dagger( z)]
                       [S(\bar x_1)-S(\bar z)]^{ab}\,\rho_p^b(\bar x_1)\right\} 
\end{eqnarray}
  whre, as before $\vec f(\vec r)=\vec r/r^2$ is the Weizsacker-Williams field of a point charge.

Eq.~(\ref{single}) is easy to interpret. The projectile appears in eq.(\ref{single}) as a distribution of the color charge density
$\rho^a_p(x)$. The charge density is a function of transverse coordinates $x$ only, since in the high energy limit all the charges are localized at $x^-=0$ and there is no nontrivial dependence on the longitudinal coordinate. The target is described as a distribution of the color fields which enter eq.(\ref{single}) through the eikonal scattering matrix of a single gluon $S^{ab}(x)$ when scattering off the target. 

For a {\it fixed} configuration of the projectile charges $\rho^a(x_1)$ in the incoming wave function, a WW gluon is emitted at a transverse position
$z$ with the probability amplitude $f(x_1-z)$. This gluon and the parent charge both scatter eikonally on a fixed configuration of the target field. The production amplitude of the gluon is determined by the eikonal amplitude $S(x_1)-S(z)$ which describes the decoherence
of the WW gluon from its emitter. The square of this amplitude gives the probability of the gluon production in the given event (for a fixed projectile and target configuration). 

This probability then has to be averaged over the projectile and target wave functions - the projectile ensemble determining the distribution of $\rho(x)$ and the (independent) target ensemble which determines the distribution of $S(x)$.
 In the framework of CGC, the averaging over the projectile and target ensembles is performed with the weight functionals
$W^P$ and $W^T$ providing the probability densities to find  a given projectile and target configurations respectively:
\beq\label{average}
\langle \hat \sigma \rangle_{P,T}\,=\,\int D\rho\, W^P[\rho_p]\, \int DS\,W^T[S]\, \hat\sigma[\rho_p,S]
\eeq
The weight functionals $W^P$ and $W^T$ are subjects to high energy evolution which determines their dependence on rapidity:
\beq
\frac{dW}{ dY}\,=\, H^{RFT}\,W.
\eeq
Here $H^{RFT}$ is the effective Hamiltonian of QCD at high energies. $H^{RFT}$ has to be derived directly from QCD and is the QCD incarnation of the Reggeon field theory of the old 
Reggeon Calculus.
The derivation of $H^{RFT}$ at the moment is incomplete. For processes involving the scattering of a dilute object on a dense one the limits of $H^{RFT}$ are known. The evolution of the wave function of a dense target which can scatter on a dilute projectile the relevant approximation is the JIMWLK evolution. The complementary form of the evolution applicable for the evolution of a dilute target, which participates in the scattering on a dense projectile,  is the so called KLWMIJ evolution derived in Ref.(\refcite{KLwmij}).
Both of these limits include a tripple-Pomeron vertex and effectively resum the
fan diagrams of the Pomeron Calculus. The JIMWLK and KLWMIJ evolutions are applicable to the same type of processes, the asymmetric situation when one of the objects is dense and the other one is dlute. The two equations describe the process in different frames: JIMWLK ascribes the evolution in energy to boosting of the dense object, while KLWMIJ to boosting of the dilute one. A more general form of $H^{RFT}$ which includes Pomeron loops is available in Ref.(\refcite{KLploop,KLploop1}), however it is not general enough to be applicable to dense-dense scattering.

Eq.~(\ref{single}) is not symmetric under the interchange of the projectile and the target, which reflects our treatment of the projectile as a dilute object. For a dense target,  %for the high energy gluon's $S$-matrix is valid,
%and the scattering process is a rotation of the color by a unitary  matrix in the adjoined representation of $SU(N_c)$ with $N_c$ being a number of colors. 
%While 
the eikonal factors $S$ are nonlinear functions of the {\it target} color charge density, which is the result of the resummation of multiple scattering on a dense target field. The projectile contribution to eq.(\ref{single}) on the other hand, is quadratic in $\rho_p$, reflecting the fact that the gluon emission amplitude in the projectile wave function is accounted for in the linear approximation valid only for small $\rho_p$.

The double-inclusive gluon production in the CGC framework has been analyzed in Ref.(\refcite{jamal,baier,KLincl,Braunincl2}). The general expression is rather long and cumbersome, but only part of it is relevant to the mechanism we discuss in this review. This is the part of the cross section where the two gluons are produced independently from two different showers. Formally this term is leading in the limit of large color  charge density of the projectile, $\rho_p\sim 1/g$.
% One should keep in mind, however, that in this limit other terms 
%not included in eq.(\ref{double}) are equally important\cite{raju1,raju2,KLploop}.
The relevant part of the double-inclusive production cross section is
\beq \label{double}
\frac{dN}{d^2pd^2kd\eta d\xi}\,=\,\langle \hat\sigma (k)\  \hat\sigma(p) \rangle_{P,T}\ + \  {\rm  terms \ subleading \ in \ \rho_p }\eeq

%As it appears, Eq.~(\ref{double})  ignores any rapidity evolution between the produced gluons.  

Eq.~(\ref{double}) does not include rapidity evolution between the rapidities of the produced gluons. It is applicable when the gluons are produced
sufficiently close to each other in rapidity so that there are not many gluons produced in the rapidity interval between them. This is not a real limitation, since the rapidity evolution implies that this is the case unless the rapidity difference between the two gluons is greater than $O(1/\alpha_s)$. For a greater rapidity separation one would have to include the evolution. The formalism for including the evolution exists\cite{jamal},\cite{KLincl} and an approximate numerical study in connection with ridge in HI has been performed\cite{DL1}, but we will not discuss this more complicated kinematics any further in this review.

It is straightforward to see that eq.(\ref{double})  indeed produces angular correlations. 
For a {\it fixed configuration} of the projectile sources $\rho_p(x)$ and target fields $S(x)$, the function $\hat\sigma(k)$ as a function of momentum has a maximum at some value ${ \vec k}={\vec Q}$.
Therefore clearly the product in eq.(\ref{double}) is maximal for ${\vec k}={ \vec p}={\vec Q}$. The value of the vector ${\vec Q}$ of course differs from one configuration to another, but the fact that momenta $\vec k$ and $\vec p$ are parallel to each other does not. Therefore after averaging over the ensemble, 
$\langle\hat \sigma(k)\,\hat \sigma(p)\rangle$ has maximum at relative zero angle between the two momenta.
The strength of the maximum of course depends on the detailed nature of the field configurations constituting the two ensembles (the projectile and the target).

We stress that even though averaged over all configuration $\langle\hat \sigma(k)\rangle _{P,T}$ must be isotropic, there is absolutely no reason for it to be isotropic for any given configuration. Thus, angular correlations emerge as a result of target/projectile averaging procedure over  isotropic ensembles  of anisotropic  
configurations\cite{KLcor}.

The simple formula eq.(\ref{double}) indeed encodes all the main features of the intuitive discussion of the previous section. The correlations originate in the initial state. They are clearly rapidity independent within the approximation in which one does not include rapidity evolution between the two observed gluons. As noted above, this is expected to be valid for rapidity difference $\eta-\xi<1/\alpha_s$.  

The correlations are peaked at the forward angle, but, consistent with our previous discussion, they also peak at the backward angle with the same strength. Clearly, 
 $\hat \sigma(k)$ is a symmetric function of its argument. It is expressed in terms of the single gluon production amplitude
 \beq
 \hat \sigma(k)=\vec a(k)\vec a^*(k)\eeq
 with
 \beq \vec a(k)=\int_{z, x_{1 },}e^{ikz} g\vec{f}( x_1- z)\, [S( x_1)-S( z)]\rho(x_1)
 \eeq
 Since the amplitude $\vec a(x)$ is real in coordinate space (eikonal scattering matrices in the eikonal approximation are by definition real), it satisfies $\vec a^*(k)=\vec a(-k)$, and therefore configuration by configuration 
\begin{equation}
\hat \sigma(k)=\hat \sigma(-k)\,.
\end{equation}           
Thus the leading contribution to the two particle inclusive production probability is symmetric 
under
\begin{equation}
\frac{dN}{d^2pd^2kd\eta d\xi}(k,p)=\frac{dN}{d^2pd^2kd\eta d\xi}(-k,p)\,. \label{symmetry}
\end{equation}
and must have two degenerate maxima, at relative angles $\Delta\phi=0,\pi$.      

Eq.(\ref{double}) also suggests that if one includes in the calculation hadrons produced from quarks, the degeneracy will be lifted. The quark production amplitude involves the eikonal matrix of a produced quark, which, being a fundamental representation matrix, is complex. Thus the $\vec k\rightarrow -\vec k$ symmetry is broken by the presence of quarks.

The overall magnitude of the correlation can be estimated from the following simple consideration. In order for the two produced gluons to be correlated in the final state, they have to be close in the initial state and also scatter off the same target field. In other words they must be within the smaller of the two correlation length that characterize the projectile and the target wave functions. The target and the projectile  correlation lengths in the saturation regime are determined by the saturation momenta $l^{-1}_{P(T)}\propto Q_s^{P(T)}$. 
 Thus for correlated production the two gluons need to be within the radius ${1/ Q_s^{max}}$ of each other, where $Q_s^{max}$ is the larger of the two saturation momenta $Q_s^P$ and $Q_s^T$. On the other hand the total number of produced gluons is proportional to the total transverse area of the smaller of the two objects participating in collision. Thus parametrically\cite{DL1,DL2}
\begin{equation}
R\,\sim    \frac{1}{(Q_s^{max})^2S_{min}}\,.
\label{estimate}
\end{equation}

Our general discussion provides an intuitive explanation to 
the correlation phenomena and also makes it clear that the presence of the correlations does not depend on the specifics of the approximation used to estimate $\frac{d^2N}{ d^2kd^2p}$. The magnitude of the effect however may depend on the approximation quite strongly. In particular what one assumes about the projectile and target averaging procedure is very important. Numerical calculations of correlations in the CGC approach have used the Gaussian weight for averaging the projectile and the target charge densities. Although the Gaussian weight makes the calculations manageable and explicit, it has certain non-generic features which most likely skew the results of the calculation. We now address this quite specific, but important point.

\section{Correlations with Gaussian averaging - the large $N_c$ conundrum.}

For the calculation of the correlated production one needs two types of averages. The simpler ones appear already in the single-inclusive gluon production. On the projectile side one needs to know
\beq\label{phi}
\langle \rho^a_p(x)\,\rho^b_p(y\rangle_P\,=\,\frac{1}{ N^2_c-1}\delta^{ab}\langle \rho^i_p(x)\rho_p^i(y)\rangle_P\, =\,\frac{1}{ N^2_c-1}\delta^{ab}\Phi^P_\eta(x,y) \ \ \ \ \ \  \ 
\eeq
$\Phi^P_\eta$ is related to  the unintegrated gluon distribution of the projectile via Fourier transform:
\beq
 \Phi^P(k_1,k_2)\,=\, \int_{x,y} e^{-ik_1x-ik_2y}\,\Phi^P(x,y) 
 \eeq
 The color structure of eq.(\ref{phi}) follows directly from the color singlet nature of the projectile wave function. Assuming translational invariance in the transverse plane, one has
 \beq\Phi^P(x,y)=\Phi^P(x-y); \ \ \ \ \ \Phi^P(k_1,k_2)= \phi^P(k_1)\,\delta^2(k_1+k_2)\eeq
 This approximation is commonly used in the literature. It should be a good approximation whenever the hadron is large enough to contain several domains of the size $Q^{-1}_S$.
%The rapidity ($Y$) dependence of the gluon distribution is determined by evolving the initial hadronic wave function with $H^{RFT}$, as mentioned above.

On the target side the relevant average is
\beq
\langle [S(x)\,S^\dagger(y)] ^{ab}\rangle_T\,=\,\frac{1}{ N^2_c-1}\delta^{ab}\langle tr [S(x)S^\dagger (y)]\rangle_T\
\eeq
Here $ tr [S(x)S^\dagger (y)]$ is the forward scattering matrix of an adjoint (gluonic) dipole. Expressed in terms of its  Fourier transform
\beq
T(k_1,k_2)\,=\,\frac{1}{ N^2_c-1}\,\int_{x,y}e^{-ik_1x-ik_2y}\,\langle tr[S(x)\,S^\dagger(y)] \rangle_T
\eeq
the single inclusive probability reads
\begin{eqnarray}\label{sing}
\frac{dN}{ d^2kd\eta }= \frac{\alpha_s}{ \pi}\int_{p,q} K_{2\rightarrow 2}(k,p,q)\,\Phi_\eta^P(p,q)\,T_{Y-\eta}(-k-p,k-q)
\end{eqnarray}
with the vertex
\beq
K_{2\rightarrow 2}(k,p,q)\,=\,\left[\frac{q_i}{ q^2}+\frac{k_i}{ k^2}\right]\,\left[\frac{k_i}{ k^2}-\frac{p_i}{ p^2}\right]
\eeq
Here $Y$ is the total rapidity interval between the target and projectile (total energy of the process) while $\eta$ is the rapidity of the observed gluon.
If either the target or the projectile are translationally invariant, Eq. (\ref{sing}) reduces to the standard $k_t$ factorized formula for DIS \cite{KT}.

The other type of averages involves the fourth power of  the gluon eikonal amplitude in the target:
\begin{equation}
\langle[S^\dagger(x)S(z)]^{ab}[S^\dagger(y)S(u)]^{cd}\rangle _T\label{4S}
\end{equation}
and similarly fourth power of the charge density in the projectile
\begin{equation}\label{4rho}
\langle\rho^a_p(x)\rho^b_p(\bar x)\rho^c_p(y)\rho^d_p(\bar y)\rangle _P\,.
\end{equation}

In the Gaussian approximation\cite{mv1,mv2} one assumes that $W^P$ (and similarly for $W^T$, see below) is a Gaussian functional in $\rho$ at any rapidity.
This leads to the factorized correlation functions:
 \begin{eqnarray}\label{rhoG}
&&\langle\rho^a_p(x)\rho^b_p(\bar x)\rho^c_p(y)\rho^d_p(\bar y)\rangle _P\,=
\left[\langle\rho^a_p(x)\rho^b_p(\bar x)\rangle_P\langle \rho^c_p(y)\rho^d_p(\bar y)\rangle _P\,+\right. \nonumber \\
&&\hspace{2cm}+\left.\langle\rho^a_p(x)\rho^c_p(y)\rangle_P\langle \rho^b_p(\bar x)\rho^d_p(\bar y)\rangle _P+
 \langle\rho^a_p(x)\rho^d_p(\bar y)\rangle_P\langle \rho^c_p(y)\rho^b_p(\bar x)\rangle _P
\right] \nonumber \\
&&\hspace{2cm}\,=\,\frac{1}{ (N^2_c-1)^2}\, \left[\delta^{ab}\delta^{cd}\Phi^P(x,\bar x)\Phi^P(y,\bar y)+\,\delta^{ac}\delta^{bd}\Phi^P(x,y)\Phi^P(\bar x,\bar y)+\right. \nonumber \\ 
&&\hspace{5cm}+\left.\delta^{ad}\delta^{bc}\Phi^P(x,\bar y)\Phi^P(y,\bar x)
\right]
\end{eqnarray} 
 %Another way of dealing with the matrix element (\ref{4rho}), which we will refer to as color singlet approximation,
Once we assume eq.(\ref{rhoG}) for the projectile averages, only three target structures enter the calculation:
\begin{eqnarray} \label{Gproduct}
&&\frac{dN}{ d^2pd^2kd\eta d\xi}\,\sim\,\Phi^P(x-\bar x)\Phi^P(y-\bar y)\,\times \nonumber \\ 
&&\times\langle tr[(S^\dagger(\bar x)-S^\dagger(\bar z))(S(x)-S( z))]
tr[(S^\dagger(y)-S^\dagger(w))(S(\bar y)-S(\bar w))]\rangle_T\,+\nonumber \\
&&+\Phi^P(x-\bar y)\Phi^P(\bar x- y)\,\times \nonumber \\ 
&&\times\langle tr[(S^\dagger(\bar x)-S^\dagger(\bar z))(S(x)-S(z)) (S^\dagger(y)-S^\dagger(w))(S(\bar y)-S(\bar w))]\rangle_T\,+\nonumber \\
&&+\Phi^P(x- y)\Phi^P(\bar x- \bar y)\,\times \nonumber \\ 
&&\times\langle tr[ (S^\dagger(y)-S^\dagger(w))(S(\bar y)-S(\bar w))(S^\dagger(\bar x)-S^\dagger(\bar z))(S(x)-S(z))]\rangle_T \nonumber \\
\end{eqnarray}

This expression contains two types of target matrix elements:   the two-dipole  matrix element 
$\langle tr[S^\dagger S]tr[S^\dagger S]\rangle_T$ and a quadrupole
$\langle tr[S^\dagger S S^\dagger S]\rangle_T$.  

In principle one is interested in dependence of the double gluon production on both, the rapidity of the observed gluons and on the total energy of the process. To study these one has to evolve the projectile averages to rapidity $\eta$ and the target averages to rapidity $Y-\eta$ with $H^{RFT}$. Current numerical implementations adopt the simplified approach, whereby the projectile averages are evolved by the BK equation, while for the target averages one
adopts an ansatze inspired by the limit of dilute target.
In the dilute target limit one can expand the eikonal factor $S$ to leading order in the target color field
% Meanwhile we want to discuss the Gaussian approximation that assumes
%that the averages factorize into products over averages. While the Gaussian approximation  for the not-so-dense  projectile is very questionable, it is definitely  more justifiable for  dense target. The question, however, is what is the quantity in the target that has a Gaussian distribution? It could be the eikonal $S$-matrix  itself.
%In such a case, a generic matrix element of four $S$ would factorize: 
%\beq \langle  S^\dagger S S^\dagger S \rangle_T\,\sim \, \langle S^\dagger\, S\rangle_T\,\langle S^\dagger\, S\rangle_T\eeq
%and the leading large $N_c$ limit is dominted by 
%  $\langle tr[S^\dagger S]\rangle_T\langle tr[S^\dagger S]\rangle_T$. Most importantly, this  term 
% does not contribute to the correlations $R$ because it exactly cancels against the uncorrelated piece.  Due to the complicated color structure mentioned above,
% it is not clear how to implement the Gaussian factorization for the quadrupole matrix element. To avoid this new problem, one can expand $S$ as 
 \beq S(x)=1+iT^a\alpha^a_T(x)\eeq
  where $\alpha^a_T$ is the color fields created by target charges, $\nabla^2\alpha_T=\rho_T$. 
%Within this expansion, the two-dipole operator corresponds 
%to an exchange of two Pomerons, while the quadrupole operator
%corresponds to the so-called cylindrical configuration of reggeized gluons. 
One further assumes Gaussian weight for the target color charge densities, or equivalently for the color fields $\alpha_T$
\begin{eqnarray}\label{alpha}
&&\langle \alpha^a_T(x)\alpha^b_T(\bar x)\alpha^c_T(y)\alpha^d_T(\bar y)\rangle_T=\frac{1}{ (N^2_c-1)^2}\times  \\
&&\times\Big[\delta^{ab}\delta^{cd}\bar \Phi^T(x,\bar x)\,\bar\Phi^T(y, \bar y)  +\delta^{ac}\delta^{bd}
\bar \Phi^T(x,y)\,\bar\Phi^T(\bar x, \bar y) +  \delta^{ad}\delta^{bc}\bar \Phi^T(x,\bar y)\,\bar\Phi^T(y, \bar x)\Big]\nonumber
\end{eqnarray}
where 
\beq
\bar \Phi^T(x,\bar x)\equiv\frac{1}{\nabla^2}(x-u)\frac{1}{\nabla^2}(\bar x-v)\Phi^T(u,v)\eeq
The rapidity dependence of the averages is then given by substituting for $\bar \Phi_{Y-\eta}$ the solution of BK equation. 

\begin{figure}[pb]\label{4g}
\vspace*{-0.2cm}
\centerline{\psfig{file=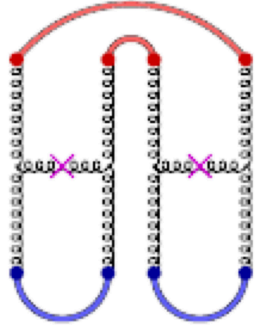,width=4cm}}
\vspace*{-0.2cm}
\caption{Correlated two-gluon production within the Gaussian approximation
%\cite{adrianjamal}
}
\end{figure}

This procedure amounts to the following ansatz for the target averages at all rapidities:
\begin{eqnarray}\label{taverage}
&&\langle(S(x)S^\dagger(y))^{ab}(S(u)S^\dagger(v))^{cd}\rangle=\delta^{ab}\delta^{cd} \frac{1}{(N^2-1)^2}\langle tr[S(x)S^\dagger(y)]\rangle \langle tr[S(u)S^\dagger(v)]\rangle\nonumber\\
&+&(T^iT^j)^{ab}(T^iT^j)^{cd}\frac{1}{N^2(N^2-1)^2}\langle tr[S(x)S^\dagger(u)-1]\rangle\langle tr[S(y)S^\dagger(v)-1]\rangle\nonumber\\
&+&(T^iT^j)^{ab}(T^jT^i)^{cd}\frac{1}{N^2(N^2-1)^2}\langle tr[S(x)S^\dagger(v)-1]\rangle \langle tr[S(u)S^\dagger(y)-1]\rangle
\end{eqnarray}
Substituting eq.(\ref{taverage}) into eq.(\ref{Gproduct}) gives an explicit expression for the double inclusive probability in terms of solutions of the BK equation.
 The structure of the final result is summarized in  a self-explanatory Fig. (\ref{4g}), 
borrowed from Ref. (\refcite{adrianjamal}).

 We will not write out all the terms explicitly, but rather stress one very important point. Namely, the correlated part of production probability in this approach is subleading in $N_c$.
This  is easy to see, since the leading $N_c$ contribution to the doubly inclusive probability {\bf in the Gaussian approximation} is given by
\begin{eqnarray}
&&\frac{dN}{ d^2pd^2kd\eta d\xi}_{Gaussian}\sim \int _{x,\bar x,y,\bar y,z,\bar z ,w,\bar w}\times\\
&\times&e^{ik(z-\bar z)}\Phi^P(x-\bar x)\langle tr[(S^\dagger(\bar x)-S^\dagger(\bar z))(S(x)-S( z))]\rangle_T\times\nonumber\\
&\times&e^{ip(w-\bar w)}\Phi^P(y-\bar y)
\langle tr[(S^\dagger(y)-S^\dagger(w))(S(\bar y)-S(\bar w))]\rangle_T\nonumber\\
&&=\frac{dN}{d^2pd\eta }\frac{dN}{d^2kd\xi}\nonumber
\end{eqnarray}
This contribution therefore cancells in the experimental ratio $R$, and only the terms suppressed by $1/N^2_c$ contribute.
This leading $N_c$ cancellation is specific to the Gaussian approximation, and is not generic. We will discuss it in more detail in the rest of this review. 

Gaussian approximation is attractive due to its relative simplicity: it allows to compute particle correlations using the known blocks computed earlier and fitted to other data. The first application of this approach gave reasonable agreement with the CMS $pp$ data\cite{ddgjlv}. More recent papers\cite{DV,DV1} include the "`nonclassical"' terms that we have not explicitly written in eq.(\ref{double}) to successfully fit $pp$  data and give a reasonable prediction for $pPb$ double production not just in the ridge kinematics but at all angles.

The theoretical status of this approach in the present context is however rather questionable. For one thing, as mentioned above we expect non-vanishing  leading $N_c$ contributions to the correlated production. The other aspect is, that even in the subleading order in $N_c$, the Gaussian approximation does not give a generic result.
In particular as was pointed out in Ref.(\refcite{adrianjamal}), the structure of the averages eq.(\ref{rhoG}) is not compatible with the BFKL and BK evolution. It was suggested in Ref. \refcite{adrianjamal} that additional terms of the form
\begin{equation}
\delta\langle \rho^a_x\rho^b_y\rho^c_u\rho^d_v\rangle=P^{ab}_{cd}[8A]M(x,y,u,v)+P^{ab}_{tcd}[8A]M(x,u,y,v)+P^{ab}_{ucd}[8A]M(x,v,u,y)
\end{equation}
with the projector $P[8A]$ defined in eq.(\ref{projectors}), should be included in the averaging procedure. One can in fact perform the complete analysis of the independent "condensates" that can appear in the projectile (target) wave function\cite{kl2012}.
 Given a color singlet projectile (target),
the most general expression for the matrix element is  expandable as a sum over
 a full set of irreducible representations to which $8\otimes 8=1+8_A+8_S+10+\bar{10}+27+R_7$ 
can be decomposed (see e.g. Ref. (\refcite{7Poms})):
\begin{eqnarray}\label{rhocolor}
\langle\rho^a_p(x)\rho^b_p(\bar x)\rho^c_p(y)\rho^d_p(\bar y)\rangle _P&=&\sum_r \frac{1}{ d_r} P_{cd }^{ab}[r] \,\Delta^P_{[r]}(x,\bar x,y,\bar y) \nonumber \\
\Delta^P_{[r]}(x,\bar x,y,\bar y)&\equiv&\langle  P_{mn}^{ij}[r]\,\rho^i(x)\rho^j(\bar x)\rho^m(y)\rho^n(\bar y)\rangle _P
\end{eqnarray}
Here $P[r]$ is a projector on representation $r$ of dimension $d[r]$ which can arise from multiplying two adjoint representations. For example
\begin{eqnarray}\label{projectors}
P[1]&=&\frac{1}{ N^2_c-1}\delta^{ab}\delta^{cd}; \ \ \ \ \ \ \ d[1]=1;\ \ \ \ \ \ \ \
P[8_A]=\frac{1}{ N_c} f^{abi}f^{icd}; \nonumber \\
P[8_S]&=&\frac{1}{ N_c} d^{abi}d^{icd};\ \ \ \ \ \ \ \ \  d[8]=N^2_c-1
\end{eqnarray}
The explicit form of the other projectors can be found in Ref.(\refcite{7Poms}).
In the dilute projectile limit the rapidity evolution of these condensates is governed by the BKP equation\cite{BKP,BKP1},
a generalized version of the BFKL  equation satisfied by reggeons containing more than two reggeized gluons. 

With a similar decomposition for the target matrix elements
\begin{eqnarray}
\langle[S^\dagger(x)S(z)]^{ab}[S^\dagger(y)S(u)]^{cd}\rangle _T&=&\sum_r \frac{1}{ d_r} P_{cd }^{ab}[r] \,\Delta^T_{[r]}(x,z,y,u)\nonumber \\
\Delta^T_{[r]}(x,z,y,u)&\equiv&
\langle  P_{mn}^{ij}[r]\,[S^\dagger(x)S(z)]^{ij}[S^\dagger(y)S(u)]^{mn}\rangle _T\
\label{4Scolor}
\end{eqnarray}
we write
\begin{eqnarray}\label{condense}
\frac{dN}{ d^2pd^2kd\eta d\xi}&\sim &\sum_r \frac{1}{ d_r}\,\Delta^P_{[r]}(x,\bar x,y,\bar y)\,\times \nonumber \\
&&[\Delta^T_{[r]}(x,\bar x,y,\bar y)       -      \Delta^T_{[r]}(x,\bar x,y,\bar w)      -    \Delta^T_{[r]}(x,\bar x,w,\bar y)      +   \Delta^T_{[r]}(x,\bar x,w,\bar w)   \nonumber \\
&-&\Delta^T_{[r]}(x,\bar z,y,\bar y)       +      \Delta^T_{[r]}(x,\bar z,y,\bar w)      +    \Delta^T_{[r]}(x,\bar z,w,\bar y)      -   \Delta^T_{[r]}(x,\bar z,w,\bar w) \nonumber \\
&-&\Delta^T_{[r]}(z,\bar x,y,\bar y)       +      \Delta^T_{[r]}(z,\bar x,y,\bar w)      +    \Delta^T_{[r]}(z,\bar x,w,\bar y)      -   \Delta^T_{[r]}(z,\bar x,w,\bar w) \nonumber \\
&+&\Delta^T_{[r]}(z,\bar z,y,\bar y)       -      \Delta^T_{[r]}(z,\bar z,y,\bar w)      -   \Delta^T_{[r]}(z,\bar z,w,\bar y)      +   \Delta^T_{[r]}(z,\bar z,w,\bar w)] \nonumber \\
\end{eqnarray}
In Eq. (\ref{condense}) we have not indicated the WW field factors, the Fourier phases and the integrals over the  transverse coordinate for the sake of brevity.  Eq.(\ref{condense}) does not presuppose Gaussian factorization, but if it is assumed, the previous results are recovered. 
The leading $N_c$ contribution comes from the singlet projector ($r=1$). The rest of the terms are $N_c$ suppressed and provide the general expression for the $1/N_c$ corrections beyond the Gaussian approximation. To calculate the condensate one would need to have a model at initial rapidity and subsequently evolve it with rapidity using $H^{RFT}$, or, if suitable, its JIMWLK limit. A model for initial conditions that goes beyond simple Gaussian was discussed in Refs.(\refcite{Baruch,Baruch1}).
 
The rapidity dependence of the condensates is a separate interesting question. The evolution of the quadrupole operator (which is a linear combination of the condensates defined above\cite{kl2012} has been a subject of active recent studies\cite{IT,IT1,DMMX}. 
It is governed by the JIMWLK equation. Given the fact that the quadrupole in $R$ emerges only at order $1/N_c^2$,  its rapidity evolution can be studied at the leading order in $N_c$. Such an evolution was  first derived in Ref. (\refcite{jamal}) 
and then later  in Refs. (\refcite{klw,KLgluon}). 

It is also important to remember,  that in order to obtain the complete $O(1/N^2_c)$ result one not only should include direct contributions  of the quadrupole and other condensates to the observable, but also take into account the $O(1/N^2_c)$ effects in the evolution of the leading two-dipole operator\cite{KKRW,AJ}. The latter involves  quadrupoles,
sextupoles\cite{kl1} and, likely also higher order multipoles given the fact that it is the adjoint dipoles that appear in the expression for $R$.

Generally speaking, beyond the Gaussian approximation, none of  the condensates $\Delta^{P,T}_{[r]}$ factorize.
In particular there is no {\it a priori} reason to assume that the factorization holds for the leading condensate
$$\Delta_{[1]}^P(x,\bar x; y,\bar y)\ne \frac{1}{ N_c^2-1}\,\Phi^P_\eta(x-\bar x)\,\Phi^P_\eta(\bar y-y)$$
Thus one does expect nonfactorizable contributions, and therefore correlated hadron production in the large $N_c$ limit.

One might think that Gaussian factorization in the large $N_c$ limit is natural due to presence of large number of degrees of freedom. However, even though the number of degrees of freedom in large $N_c$ limit is large, the theory has legitimate states which contain small number of particles. A color dipole is an example of such state. It is a superposition of many states (different color orientations) of two particles, rather than a state with many particles. This type of state maintains high level of color coherence in the large $N_c$ limit, and is not subject to the central limit theorem. Fluctuations in density in a state like this can be large even at large $N_c$, and large fluctuations in the ensemble break factorization of correlation functions. 

A similar question was considered a while ago in Refs.(\refcite{yoshi,yoshi1}) in connection with factorization of dipole densities in the dipole model\cite{dipole,dipole1,dipole2}.
As shown in Refs.(\refcite{yoshi,yoshi1}) within the dipole model (which is defined entirely within the  large $N_c$ limit\cite{kl1}) the product of two densities does not factorize. When evolved from a single dipole initial state, the connected correlator of dipole densities of small dipoles (of size $r$, much smaller than that of the initial dipole $r_0$) as a function of the distance $b\gg r$ between the two, behave as
\beq
\langle n(r,0)n(r,b)\rangle-\langle n(r,0)\rangle\langle n(r,b)\rangle=\langle n(r,0)\rangle\langle n(r,b)\rangle \left(\frac{b}{r_0}\right)^{-\lambda}
\eeq
where $\lambda>0$ is a number. This is not quite the regime of interest to us, since we need to consider the situation when both dipoles are at the same impact parameter. Refs.(\refcite{yoshi,yoshi1}) also did not consider angular correlations. However it demonstrates as a matter of principle, that large $N_c$ limit does not eliminate correlations. 

Gaussian averaging procedure described above is much more restrictive than large $N_c$ limit, and it is very likely  that it tends to underestimate correlations.

\section{The dipole model at large $N_c$.} 

The discussion in the rest of this review will be devoted to attempts to understand the correlations which are leading in the large $N_c$ limit.
The leading $N_c$ contribution  is proportional to 
$$\Delta^P_{[1]}\,\langle tr[S(x_1)\,S^\dagger(x_2)]\,tr[S(x_3) \,S^\dagger(x_4)]\rangle_T$$

As discussed in detail above, the region of the phase space responsible for correlations is when all four points $x_i$ lie within the correlation length  $1/Q_s$ of each other. In this regime one does not expect Gaussian like factorization of averages. Thus one needs a model of the target that encodes such correlations. It is also important to understand how the correlations evolve with energy. The purpose of this section is to describe an initial study of these questions performed in 
Ref.(\refcite{KLang}). 

We study the rapidity evolution of an initial ensemble which encodes nontrivial correlation, within the projectile dipole model. This model is a leading $N_c$ approximation, and therefore fits our purpose well. Although the actual observable of interest is the correlated gluon production, which is directly related to the average of two adjoint dipoles, for the sake of simplicity we will study correlations of fundamental dipoles. The fundamental dipoles $s(x,y)$ are the basic degrees of freedom of the dipole model.
%The target probability distribution  $W^T[s]$  in the projectile dipole model evolves with rapidity according to \cite{LL,kl1}
%\begin{equation}
 %{d\over d Y}W^T[s]=
 %frac{ \bar{\alpha}_s}{2\,\pi}\,
%\int_{x,y,z}\frac{(x-y)^2}{(x-z)^2\,(z\,-\,y)^2}\,\left[\,s(x,\,y)\,-\,
%\,s(x, z)\,s(y,z)\,\,\right]
%\frac{\delta}{\delta s(x, y)} W^T[s]
%\end{equation}

Our strategy is to choose an ensemble $W^T_0[s]$ of initial configurations $s(x,y)$,  which contains nontrivial angular correlations.  Each configuration of the ensemble is evolved independently according to the BK equation\cite{Kovchegov}
\begin{equation}\label{BK}
 \frac{ds(x,y)}{ d Y}=
 \frac{ \bar{\alpha}_s}{2\,\pi}\,
\int_{z}\frac{(x-y)^2}{(x-z)^2\,(z\,-\,y)^2}\,\left[\,s(x,\,y)\,-\,\,s(x, z)\,s(y,z)\,\,\right]
\end{equation}
with $\bar \alpha$ - the 'tHooft coupling, which is finite at infinite $N_c$.
The correlations at the final rapidity are then calculated by averaging the correlator over the ensemble of solutions $s_Y(x,y) $\cite{kl1}:
\begin{equation}\label{av}
\int Ds W_Y^T[s]s(x,y)s(u,v)= \int Ds W_0^T[s]s_Y(x,y)s_Y(u,v)
\end{equation}
where $s_Y(x,y)$ is the solution of the BK equation (\ref{BK}) with initial condition $s(x,y)$.
In order to study  angular correlations we have to allow the individual members of the initial ensemble to be anisotropic. 
The rotational invariance is then restored by averaging over the whole ensemble and not configuration by configuration. 
%\beq\label{BK}
%\partial_Y\,N(\vec r)\,=\, \frac{ C_F\,{\alpha}_s}{2\,\pi}\,\int d^2 \vec r^\prime {\vec r^2\over  \vec r^{\prime\,2}\, (\vec r-\vec r)^2}\ [N(\vec r^\prime)\,+\,N(\vec r-\vec r^\prime) \,-\, N(\vec r)\,-\,N(r^\prime)\,N(\vec r-\vec r^\prime)]\,.
%\eeq
 In Eq.(\ref{BK}) $s(x,y)$ is the dipole scattering amplitude for a single configuration of the target field, rather than a target average.

Like in most BK studies, we take the target to be translationally invariant, configuration by configuration.  That is, we take the dipole scattering amplitude for all configurations of the ensemble not to depend on impact parameter, $s(x,y)=s(\vec r)$,  with $\vec r=\vec x-\vec y$.
The reason for this choice is solely the simplicity of implementation. One certainly should study the impact parameter dependent configurations. In the context of the present work we view the impact parameter independence as a proxy to studying dipoles that scatter at the same impact parameter. Intuitively one expects impact parameter variation not to change the behavior of dipoles scattering at the same impact parameter too strongly. This question however has to be studied separately.

We start the evolution with a simple  one parameter ensemble at some initial rapidity $Y_0$: 
\beq\label{ini}
s(Y_0,\vec r)\,=\,exp\{\,-\,(\vec r \vec E)^2\}\,=\,exp\{\,-\,r^2 \, E^2\,cos^2(\theta-\delta)\};\
\eeq
The parameter $\vec E$ represents the chromo electric field in a given target configuration. The field $\vec E$ is not colored and thus cannot be associated directly with the color field. Instead a better way to think about it is as representing color averaged moments over the (color invariant) target ensemble, e.g. $E_iE_j\sim <E^a_iE_j^a>$.   The magnitude of $E$ plays the very same role as $Q_s$.

Here $\theta$ and $\delta$ are the angles of the dipole moment and the field relative to an arbitrary axis. 
The absolute value of the field $E$ was fixed in this study. The angle $\delta$  was taken to be a random variable, so that the weight function for the target ensemble is a constant,  $W^T_{Y0}(\delta)=1/2\pi$.

The ensemble we work with, albeit very simple, differs significantly from previous numerous studies of the BK equation. In these studies the initial conditions are parameterized as a function of the dipole size only, without the angular dependence. This includes papers which do allow for fluctuations in the initial ensemble. For example Refs.(\refcite{AM,AA}) allow for the fluctuations of the strength of the field $E$ but without angular dependence. Such an ensemble, of course, is not suitable for a study of angular correlations. Eq.(\ref{ini}) on the other hand provides a two-dimensional initial data set. 

\begin{figure}[pb]
 \begin{tabular} {cc}
 \psfig{file=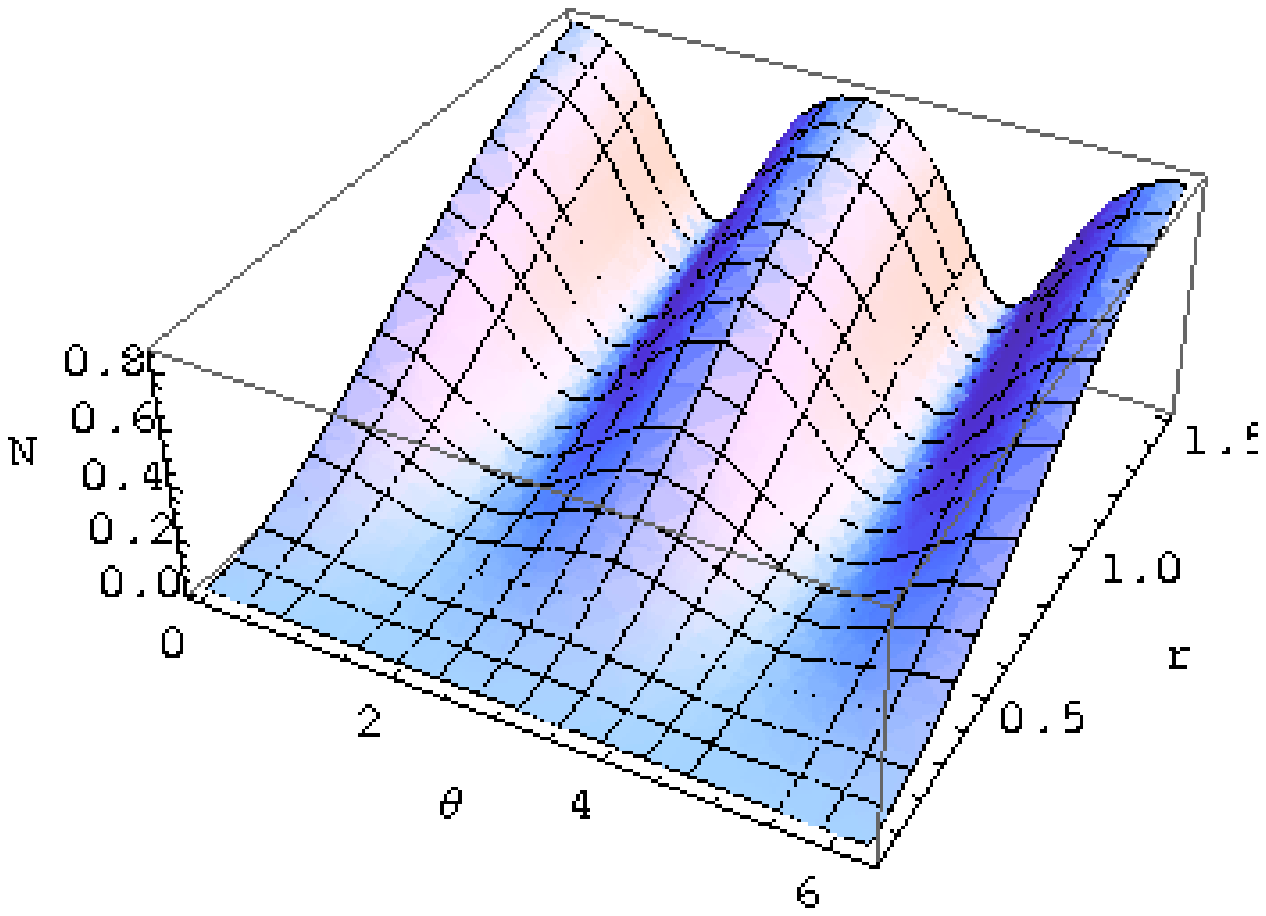,width=6cm}\ \ \ \ \ \ \  \ & \ \ \ \ \ \ \ 
\psfig{file=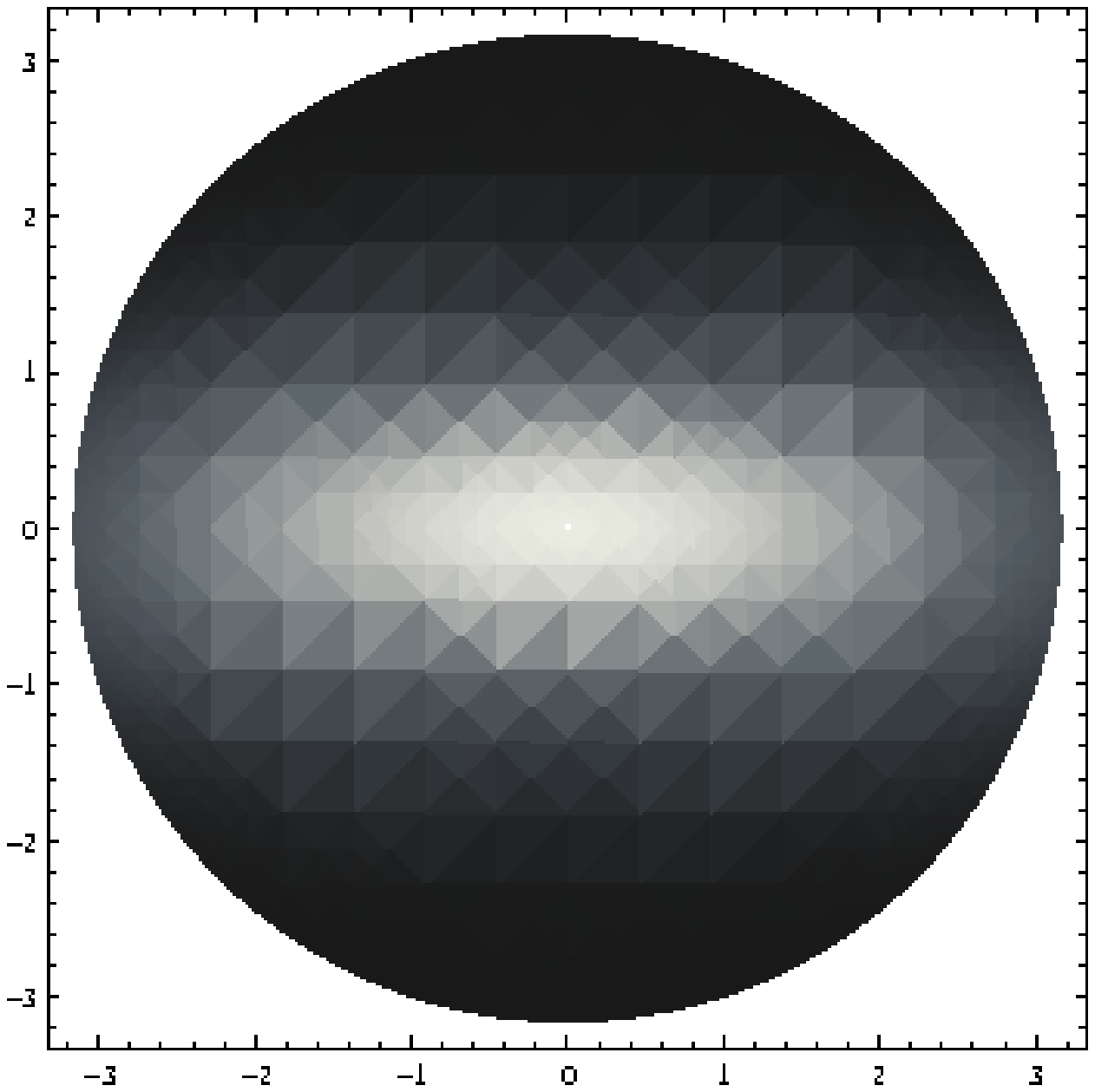,width=4cm}\ \
\end{tabular}
\caption{ Initial conditions.   Left: profile of $N$ as a function of $r$ and $\theta$. Right: the same in polar coordinates; the "blackness" of the target
is shown with respect to the dipole's orientation. \label{inifig}}
\end{figure}

Fig. (\ref{inifig}) displays the initial condition plotted for the imaginary part of the dipole scattering amplitude $N(\vec r,\delta=0) \equiv 1-s(\vec r)$. 
It is quite obvious that the initial conditions (\ref{ini}) with the uniform distribution of $\delta$ generate angular correlations between two dipoles. In particular for small dipoles ($r, r'\ll Q_s^{-1}$)
$$
\langle s(\vec r)\,s(\vec r^\prime)\rangle_{\delta}\,\sim\, r^2r'^2+2(\vec r \cdot\vec r^\prime)^2\,.
$$

Each configuration of the ensemble is evolved independently according to the BK equation. 
The correlations at the final rapidity are then calculated by averaging the correlator over the ensemble of initial conditions according to (\ref{av}). 

 We observed in Ref.(\refcite{KLang}) 
 that for any given initial condition, the angular dependence of $N(\vec r)$ is washed away  exponentially quickly with rapidity. As a consequence,
the initial correlations also quickly disappear with rapidity. In the range of rapidities studied in Ref.(\refcite{KLang}) this fall off is approximately exponential
\beq\label{expfall}
R\,\sim\,e^{-\lambda Y} \,,\ \ \ \ \ \ \ \ \ \ \ \lambda\simeq 0.6
\eeq
This result is somewhat surprising in view of our previous arguments. We would expect a similar fall off for large dipole sizes, namely those sizes that cross the saturation boundary during the evolution $r>Q_s^{-1}(Y)$, but not for small size dipoles which stay inside the saturation radius for all rapidities studied.
The fall off observed in Ref.(\refcite{KLang}) on the other hand, is rather uniform in dipole size and takes place also for very small dipoles.

Thus it appears that, even if the initial ensemble contains nontrivial leading $N_c$ correlations, after a short rapidity evolution the correlations disappear. 
This is  consistent with recent numerical simulations and analytic ansatze for solutions of the JIMWLK equation.  The solution seems to converge to ensembles with pretty much  Gaussians  properties\cite{JIMWLKnum,JIMWLKnum1,IT,IT1}.  

At first sight this is inconsistent with the results of Refs.(\refcite{yoshi,yoshi1}), as quoted above.
However the physics of this decrease, as elucidated in Ref.(\refcite{KLang}), is fairly simple and is not in contradiction with Refs.(\refcite{yoshi,yoshi1}).
The calculation of Ref.(\refcite{KLang}) is performed in the framework of the projectile dipole model, while that of Refs.(\refcite{yoshi,yoshi1}) in the target dipole model. 
Technically the projectile dipole model assumes that the projectile is dilute and contains a small number of dipoles, while the target is a dense system, while the target dipole model reverses the roles of the two objects. In terms of the target evolution, the projectile dipole model corresponds to the large $N_c$ limit of the JIMWLK evolution, while the target dipole model to that of KLWMIJ.

As discussed in Ref.(\refcite{bubbles}) the JIMWLK and KLWMIJ evolutions are very different as far as the target wave function is concerned. KLWMIJ is the normal perturbative evolution of the target wave function. It is well known that in such evolution the number of gluons grows exponentially fast. The target gluon density, which in this regime satisfies the BFKL equation, depends on the total rapidity as
\begin{equation}
g(p,Y)\propto e^{c\alpha_s Y}
\end{equation}
where $c$ is a number of order one and $Y$ is the total rapidity by which the target wave function has been boosted from "rest". The exponential dependence also holds for the differential gluon density in the wave function at any rapidity $\eta<Y$, namely (we suppress the momentum dependence in the prefactor for simplicity)
\begin{equation}
\frac{d}{d\eta}g(p,\eta)\propto e^{c\alpha_s \eta}\theta(Y-\eta)\,.
\end{equation}
Gluons in the wave function which are separated in rapidity by no more than $\delta\eta\sim O(1/\alpha_s)$ are expected to be correlated. Since the gluon density in the wave function exponentially grows with rapidity, any projectile that probes such a wave function effectively feels only the gluons in the last rapidity "bin". Thus such a probe is sensitive to any correlation that exists between the softest gluons in the wave function.

The rapidity dependence of gluon density generated by the JIMWLK evolution on the other hand, is very different. As discussed in Ref.(\refcite{bubbles}), while the probability to emit an additional gluon in the dilute regime is proportional to the number of gluons in the wave function, in the dense regime this probability approaches a constant. The evolution with constant probability of emission generates gluon density which is uniformly distributed in rapidity. This can be thought of as a random walk in color space as in Ref.(\refcite{bubbles}). So now one has 
\begin{equation}
\frac{d}{d\eta}g(p,\eta)= C\,.
\end{equation}
with $C$ a function of transverse momentum, but not of rapidity.
It should still be true, like in the KLWMIJ case, that gluons separated by a small rapidity interval are correlated, while the correlation disappears for gluons at very different rapidities. However now, as opposed to KLWMIJ case, if one scatters any projectile on such a target, the projectile will sample gluons at all rapidities equally. Thus, for example, if the projectile consists of two dipoles, the two dipoles will most likely scatter on color field components (gluons) with very different rapidities. Since such fields are not correlated, the two dipoles will scatter independently and the two dipole scattering amplitude will not exhibit any correlations.

Another way of understanding the difference between the nature of JIMWLK and KLWMIJ evolutions is the following. In KLWMIJ evolution one starts initially with the target wave function which contains a small number of gluons. The probability to emit an additional gluon in one step of the evolution is small ($O(\alpha_s)$), however if a gluon is emitted, the end configuration strongly differs from the initial one, since the number of gluons have changed by a factor of order unity. Such evolution thus generates a very rough ensemble of target field (gluon) configurations. In this ensemble configurations with very different properties are present, albeit with small weight. Such an ensemble exhibits large fluctuations in a variety of observables. 

On the other hand, in JIMWLK evolution the probability to emit an extra gluon is large - of order unity. However emission of an extra gluon produces a new configuration which is hardly distinct from the existing one, since the one extra gluon is produced on the background of $\frac{1}{\alpha_s}$ gluons which already exist in the wave function. Thus the JIMWLK ensemble  contains many configurations, but all these configurations have very similar properties. The fluctuations in observables in such an ensemble are very small. 

Thus it is reasonable to expect that KLWMIJ evolution produces (and preserves) correlations (and fluctuations) between scattering amplitudes of two projectile dipoles at leading order in $N_c$. These are indeed the correlations discussed in Ref.(\refcite{yoshi,yoshi1}). On the other hand JIMWLK evolution must lead to disappearance of any correlations initially present in the target ensemble, and this is exhibited by the results of Ref.(\refcite{KLang}).

Given that the two approximation to high energy evolution lead to such qualitatively different answers for the observable we are interested in, naturally one should ask which one of them, if any, should be used for quantitative investigations. Naively one might think, that since we are interested in high multiplicity situation, the target is dense and JIMWLK evolution is more suitable. This however is not the case. We expect the correlations to arise due to scattering at close values of impact parameter. The transverse distances in question should be smaller or of the order of the saturation radius $Q^{-1}_s$. This means that the scattering occurs on the components of the target color field with transverse momenta $p>Q_s$. However at these momenta the target wave function by definition is still dilute.  Even for a dense target JIMWLK evolution is not appropriate for all wavelengths, but only for those greater than the saturation radius.
At large momenta $p>Q_s$ the appropriate evolution is KLWMIJ and not JIMWLK.

Thus to be able to describe fluctuations one needs to be able to evolve the high momentum modes according to KLWMIJ while low momentum modes according to JIMWLK. In fact, the situation is even more complicated. One expects that the largest contribution to correlations comes from the modes with $p$ of order $Q_s$, since in the relative distances of order $Q^{-1}_s$ should dominate the integral over impact parameter. For these modes however, the Pomeron merging and splitting (JIMWLK and KLWMIJ contributions) are of equal importance. In other words, this momentum range is sensitive at the leading order to the Pomeron loop contributions. We thus conclude that Pomeron loops need to be included for proper treatment of this question.

%Comparing Refs. (\refcite{LR}) and (\refcite{KLang})  we notice that the way the correlations are introduced in the initial conditions are essentially  equivalent.
%However the rapidity evolution is implemented quite differently in the two approaches. While the latter  traces the two-dipole evolution exactly as follows
%from the dipole model, the former postulates the ansatze (\ref{Q0}) at any value of rapidity. As a result, the rapidity dependence in the two approach came out
%quite different: the exponential fall off in Ref.  (\refcite{KLang}) vs a mild dependence in  (\refcite{LR}).
 
%Thus, putting aside  yet fully unexplored effects of  pomeron loops,  
%the dominant correlations should be searched  at $1/N_c^2$ level. 

The impact parameter independence of the distribution considered in Ref.(\refcite{KLang}) obviously misses an important part of dynamics, as it largely based on hot spot picture of the transverse plane.
The full picture should include the electric field $\vec E=\vec E(\vec b)$
 fluctuating both in magnitude and direction throughout the transverse plane (Fig. \ref{domain}).  This study is currently in progress\cite{Andrej}.

\section{Double Pomeron exchange model}
As mentioned in the introduction, there is another approach to angular correlations in the recent literature\cite{LR}, that seems to be rather close to the one discussed here. In this section we wish to discuss it somewhat further, and also point out that superficial similarities notwithstanding, the physics described by the approach of Ref.(\refcite{LR}) is quite different. We will not follow Ref.(\refcite{LR}) verbatim, but rather will give our interpretation of the approach presented there.

Let us examine more carefully our original expression eq.(\ref{double}). For the purpose of the present argument we will assume that the production is dominated by the part of the phase space where the charge densities $\rho(x_i)$ are far enough from the emitted gluons, so that the gluons necessarily decohere and materialize in the final state once they scatter.
This is essentially partonic picture of the projectile, where the gluons in the incoming wave function are assumed to carry only very small transverse momentum, and all the momentum in the final state comes from the exchange with the target\cite{tolgaproduction}. Although this contribution does not account for all the gluon production, it is likely to dominate the correlated production at $\phi\sim 0$. With this assumption we can simplify our expressions by dropping the eikonal factors associated with the scattering of the valence color charge densities.
The expression for double inclusive gluon production then becomes
\begin{eqnarray}\label{qne0}
&&\frac{dN}{ d^2pd^2kd\eta d\xi}\,=\,\langle \hat\sigma (k)\  \hat\sigma(p)\rangle_{P,T}
\,=\,\frac{\alpha^2}{\pi^2}\int \frac{d^2l}{(2\pi)^2}\frac{d^2\bar l}{(2\pi)^2}\frac{d^2Q}{(2\pi)^2} \times \nonumber \\
&&\times\left[\left(\frac{k_i}{k^2}-\frac{l_i-Q_i}{(l-Q)^2}\right)\left(\frac{k_i}{k^2}-\frac{l_i+Q_i}{(l+Q)^2}\right)\right]
\left[\left(\frac{p_j}{p^2}-\frac{\bar l_j+Q_j}{(\bar l+Q)^2}\right)\left(\frac{p_i}{p^2}-\frac{\bar l_j-Q_i}{(\bar l-Q)^2}\right)\right]\nonumber\\
&&\nonumber \\
&&\times\langle \hat n(l,Q)\hat n(\bar l, -Q)\rangle_P\langle \hat T(k-l,-Q)\hat T(p-\bar l, Q)\rangle_T
\end{eqnarray}
with 
\begin{eqnarray}
\hat n(l,Q)&\equiv& \int d^2xd^2y e^{il(x-y)+iQ(x+y)}\rho^a(x)\rho^a(y); \nonumber \\
\hat T(k,Q)&\equiv& \int d^2xd^2y e^{ik(x-y)+iQ(x+y)}{\rm Tr}[S(x)S^\dagger(y)]
\end{eqnarray}
Here $\hat n$ ($\hat T$) are operators on the projectile (target) Hilbert spaces such that
\begin{eqnarray}
 \Phi^P(-l-Q,-l+Q)&=&\langle \hat n(l,Q) \rangle_P; \nonumber \\
T(-l-Q,-l+Q)&=& {1\over N_c^2-1}\,\langle \hat T(l,Q) \rangle_T
\end{eqnarray}

In the above equations we have assumed that the projectile and the target are both described by translationally invariant ensembles. Although we have not indicated this explicitly, both $\hat n$ and $\hat T$ depend on rapidity, so that strictly speaking one should write $\hat n_\eta$, $\hat T_{Y-\eta}$. Note, that the physical meaning of $Q$ is not that of momentum transferred in the scattering, but rather that of the initial transverse momentum of the dipole with fixed size $\vec r$. Thus $\hat T(\vec Q)$ has the meaning of {\bf forward} scattering amplitude of the dipole with transverse momentum $\vec Q$. 

For single inclusive gluon emission amplitude we again quote eq. (\ref{single})
\begin{equation}\label{sn}
\frac{dN}{ d^2kd\eta }=\frac{\alpha}{\pi}\int \frac{d^2l}{(2\pi)^2}\left[\left(\frac{k_i}{k^2}-\frac{l_i}{l^2}\right)\left(\frac{k_i}{k^2}-\frac{l_i}{l^2}\right)\right]\langle n(l,Q=0)\rangle_P\langle T(k-l,-Q=0)\rangle_T
\end{equation}             
The double inclusive cross section can be written in terms of a natural generalization of eq.(\ref{sn}). With minor abuse of notations we introduce an operator
\begin{equation}\label{nq}
\frac{d\hat N(Q)}{ d^2kd\eta }\equiv\frac{\alpha}{\pi}\int \frac{d^2l}{(2\pi)^2}\left[\left(\frac{k_i}{k^2}-\frac{l_i-Q_i}{(l-Q)^2}\right)\left(\frac{k_i}{k^2}-\frac{l_i+Q_i}{(l+Q)^2}\right)\right]\hat n(l,Q)\hat T(k-l,-Q)
\end{equation}
keeping in mind that eq.(\ref{nq}) is defined configuration by configuration for fixed values of projectile density and target field. This quantity (up to kinematical factors) can be though of as a Pomeron extending from the projectile to target rapidity which is cut at the rapidity where the final state gluon is measured. 

The translational invariance of the target and projectile therefore restricts the contributions to the single inclusive cross section to zero momentum 
transfer Pomeron. For double inclusive gluon however, only the total momentum transfer between the two Pomerons has to vanish, while each Pomeron individually can transfer a nonvanishing momentum $Q$. Thus {\it a priori}, there are two distinct sources of nonfactorizability of the double inclusive cross section, and thus of correlations. First, is possible nonfactorizability of the $Q=0$ amplitudes, and second direct contribution of the $Q\ne 0$ terms.

In the calculation described in the previous section\cite{KLang}, by limiting ourselves to the target ensemble translationally invariant {\it configuration by configuration} we have limited ourselves to the $Q=0$ contribution, and found that the initial correlations fall off exponentially with rapidity.  On the other hand Ref.(\refcite{LR}) considers the effect of $Q\ne 0$ contributions within the framework of the BFKL evolution and finds correlations which survive at arbitrary rapidity.

\subsection{Disappearance of correlations at $Q=0$.}

Within the BFKL framework at leading $N_c$ initial correlations possibly present in the $Q=0$ channel, disappear with rapidity. Consider for definiteness the target averages.
In the large $N_c$ limit, the solution of the BFKL (or rather BKP) equation for $\langle T(k,Q)T(p,-Q)\rangle_T$ 
can be written in terms of the eigenfunctions of the BFKL equation for the single Pomeron.
Recall that the complete basis of solutions of the BFKL equation is \cite{Lipatov1}
\beq\label{enn}
\psi_{n,\nu,\rho}(x,y)\,= \, \left( \frac{(x-y)}{(x-\rho) \,
(y-\rho)} \right)^{\frac{1+n}{2} + i \nu} \ \left(
\frac{(x-y)^*}{(x-\rho)^* \, (y-\rho)^*} \right)^{\frac{1-n}{2} + i
\nu}
\eeq
where the complex coordinate $x$ is defined as $x=x_1+ix_2$
The corresponding eigenvalues do not depend on $\rho$
\beq\label{eig}
\omega_{n,\nu}\,=\,2 \, \bar\alpha_s \, \chi (n, \nu),
\eeq
where
\beq\label{chi}
\chi (n, \nu) \, = \, \psi (1) - \frac{1}{2} \, \psi \left(
\frac{1+|n|}{2} + i \nu \right)  - \frac{1}{2} \, \psi \left(
\frac{1+|n|}{2} - i \nu \right),
\eeq
These wave functions  satisfy the completeness relation
\begin{eqnarray}\label{EE}
(2 \pi)^4 \, \delta (x-\bar x) \, \delta (y-\bar y) &= &
\sum_{n=-\infty}^\infty \int_{-\infty}^\infty d \nu \int d^2
\rho \, \frac{16 \left(\nu^2 + \frac{n^2}{4}\right)}{|x-y|^2 
|\bar x-\bar y|^2} \, \times \nonumber \\ &&  \nonumber \\  
&\times& \psi^{n, \nu} (x-\rho, y-\rho)
\psi^{n, \nu \, *} (\bar x-\rho, \bar y-\rho)
\end{eqnarray}      
The momentum space representation of the BFKL eigenfunctions 
\begin{equation}
\psi_{n,\nu,\rho}(k,Q)=\int d^2(x-y) \int d^2(x+y)e^{ik\cdot(x-y)+iQ\cdot\frac{(x+y)}{2}}\psi_{n,\nu,\rho}(x,y).
\end{equation}
For translationally invariant solutions at $Q=0$ the basis is
\begin{equation}
\psi_{n,\nu}(k)= \int d^2(x-y) e^{ik(x-y)}\int d^2\rho \ \psi_{n,\nu,\rho}(x,y)
\end{equation}  
At large $N_c$ we have
\begin{eqnarray}\label{rotat}
\langle T(k,Q=0)T(p,Q=0)\rangle_T&=&\int d\nu_!\,d\nu_2\,\Sigma_n\Phi(\nu_1,\nu_2,n)\psi_{n,\nu_1}(k)\psi_{-n,\nu_2}(p)\\
&\approx& \Sigma_n\Phi(n)\psi_{n}(k)\psi_{-n}(p)\nonumber
\end{eqnarray}
where we have assumed rotational invariance of the target ensemble. We have also assumed for simplicity that the integral over $\nu_{1,2}$ is dominated by the saddle point $\nu_{1,2}=0$, since this corresponds to highest eigenvalue at every $n$. This assumption is not strictly valid for arbitrary transverse momenta. In particular one expects that at high $k,p$ the saddle point will depend on the values of momenta\cite{yoshi}, but we do not expect this dependence to change our argument.
Due to the presence of $n\ne 0$ terms this sum contains angular correlations between the two densities. At the initial rapidity these correlations, as encoded in our initial ensemble determine the coefficient function $\Phi(n\ne 0)$. In particular, the initial distribution eq.(\ref{ini}) contains all even angular momenta $n$.
Within BFKL evolution each term in the sum in eq.(\ref{rotat}) evolves with rapidity with its own exponential
\begin{equation}
\Phi_Y(n)\propto e^{2\omega_n(Y-Y_0)}\Phi_{Y_0}(n)
\end{equation}
The eigenvalue $\omega_0$ is the standard BFKL intercept and is positive, while all other eigenvalues are negative. 
For $\nu=0$ 
\begin{equation}
\omega_{n=0}=4\ln 2\bar\alpha_s; \ \ \ \ \ \ \ \omega_{n=2}=4(\ln 2-1)\bar\alpha_s
\end{equation}
Thus the $n\ne 0$ contributions decay exponentially and
after short rapidity interval the $n=0$ component totally dominates the sum
\begin{equation}\label{ty}
\langle \hat T_Y(k,Q=0)\hat T_Y(p,Q=0)\rangle_T\rightarrow_{Y\rightarrow\infty}\ \sim \psi^Y_{0}(k)\psi^Y_{0}(p)
\end{equation}
For $n=0$ the function $\psi(k)$ does not depend on the direction of momentum $k$ and therefore eq.(\ref{ty}) does not contain any angular correlations. 
The result quoted in (\ref{expfall}) can be indeed identified with  $\omega_{n=2}$ evaluated at a reasonable value of $\bar\alpha_s$.
This is indeed the origin of the exponential disappearance of angular correlations in Ref. (\refcite{KLang}).

\subsection{The $Q\ne 0$ contribution.}

Ref.(\refcite{LR}) considers the general case of $Q\ne 0$ and $n=0$. In this case one has
\begin{eqnarray}\label{tq}
\langle \hat T(k,Q)\hat T(p,-Q)\rangle_{n=0}&=&\int d^2\rho_1\int d^2\rho_2\int_\nu \Phi_{\nu_1,\nu_2}(\rho_1,\rho_2)\psi_{\nu_1,\rho_1}(k,Q)\psi_{\nu_2,\rho_2}(p,-Q)\nonumber\\
&\approx&\int d^2\rho_1\int d^2\rho_2 \Phi(\rho_1,\rho_2)\psi_{\rho_1}(k,Q)\psi_{\rho_2}(p,-Q)
\end{eqnarray}
where
\begin{equation}
\psi_{\nu,\rho}(k,Q)=\int d^2(x-y) \int d^2(x+y)e^{ik\cdot(x-y)+iQ\cdot\frac{(x+y)}{2}}\psi_{n=0,\nu,\rho}(x,y)
\end{equation}
and in the second line of (\ref{tq}) we have again assumed that the $\nu$ integral is dominated by the saddle point $\nu_{1,2}=0$ ($\psi_{\rho}\equiv\psi_{\nu=0,\rho}$).
In general eq.(\ref{tq}) contains angular correlations. Since $\psi$ depends on the impact parameter $b\equiv x+y$ only in combination with $\rho$, one can shift $b$ by $\rho$ in the Fourier integral to obtain
\begin{eqnarray}\label{tnn0}
\langle \hat T(k,Q)\hat T(p,-Q)\rangle_{n=0}&=&\int d^2\rho_1 d^2\rho_2 e^{iQ\cdot(\rho_1-\rho_2)}\Phi(\rho_1,\rho_2)
\psi_{\rho=0}( k, Q)\psi_{\rho=0}(p,-  Q)\nonumber\\
&=&\Phi(Q)\psi_{\rho=0}( k,Q)\psi_{\rho=0}( p, -Q)
\end{eqnarray}
Note that 
\beq
\psi_{\rho=0}( k, Q)=\psi(k^2,Q^2,\vec k\cdot \vec Q)
\eeq
This expression correlates the direction of $\vec k$ with that of $\vec p$, through the dependence on $\vec Q$. This is true even for the rotationally invariant target ensemble, which in the present context means $\Phi(\vec Q)=\Phi(Q^2)$.

In particuar Ref.(\refcite{LR}) assumes for illustrative purposes
 \beq\label{Q0}
 \frac{d\hat N(Q)}{dk^2d\eta}-\frac{d\hat N(Q=0)}{dk^2d\eta}\sim (\vec k\cdot \vec Q)^2\,\ \frac{d\bar N}{dk^2d\eta}
 \eeq
 Then the correlated part of the double gluon production is 
\begin{eqnarray}\label{levin}
 \frac{d^2N}{dk^2d\eta d^2pd\xi}_{correlated}&=&\int d^2Q\,\Phi(Q^2)\,\,(\vec k\cdot \vec Q)^2\,
(\vec p\cdot \vec Q)^2\langle\frac{d\bar N}{dk^2d\eta}
 \frac{d\bar N}{dp^2d\xi}\rangle_{P,T}
 \sim\nonumber \\
&\sim& (\vec k\cdot \vec p)^2\,\langle\frac{d\bar N}{dk^2d\eta}
 \frac{d\bar N}{dp^2d\xi}\rangle_{P,T}
 \end{eqnarray}

Thus one finds correlation, which conforming to the earlier discussion has a peak when the two momenta are either parallel or antiparallel.
Since the expression in eq.(\ref{tnn0}) involves only the BFKL eigenfunctions with $n=0$, it has the leading energy dependence and in the dilute regime grows with energy with the double Pomeron intercept. Thus the angular correlations encoded in eq.(\ref{tnn0}) survive at high energy.

So it looks like the result of Ref.(\refcite{LR}) is very much in line with the qualitative picture we outlined in the first sections of this review. However, the situation is not quite that simple.  Consider a single Pomeron amplitude at fixed impact parameter
\begin{equation}\label{t00}
\hat T(\vec r,\vec b)=\int d^2\rho P(\vec \rho)\frac {r^2}{|\vec r-\vec b-\vec\rho||\vec r+\vec b+\vec\rho|}=\int d^2\rho P(\vec\rho-\vec b)\frac {r^2}{|\vec r-\vec\rho||\vec r+\vec\rho|}
\end{equation}
Now suppose that the function $P$ has a maximum for some value of its argument, and for the sake of definiteness let us choose this value zero. Then the integral is dominated by $\vec\rho=\vec b$, and we get
\begin{equation}
\hat T(\vec r,\vec b)\propto\frac {r^2}{|\vec r-\vec b||\vec r+\vec b|}
\end{equation}
For small values of the dipole size (smaller than the impact parameter $b$) 
\begin{equation}
\hat T(\vec r,\vec b)\propto\frac{r^2}{b^2}
\end{equation}
This is just the rotationally invariant scattering amplitude of a small dipole on a density profile that decreases as $1/b^2$. Intuitively a dipole will scatter this way on an ensemble of fields, whose magnitude decreases according to Coulomb law away from a central charge at $b=0$. The ensemble itself is 
rotationally invariant, which leads to an isotropic scattering amplitude.
On the other hand for $r^2\gg b^2$ we have
\begin{equation}
\hat T(\vec r,\vec b)=const
\end{equation}
So a dipole scattering in the center of the distribution again has an isotropic scattering amplitude. 
There is an orientation dependence only  for dipoles of the size comparable with the impact parameter $b$. 

This suggests the following picture of the "target" described by the
amplitude eq.(\ref{t00}). It is a position dependent distribution with a center at $\vec x=0$. The distribution itself is locally isotropic. This means that it is not anything like a single electric field configuration, which has locally a direction in space, but rather a locally isotropic ensemble of such field configurations. The mean density of this ensemble decreases as $1/b^2$ away from the center. A small size dipole scattering on periphery does not care about orientation, since it again sees a locally isotropic distribution. A dipole whose size is of the same order as a the impact parameter probes different densities depending on its orientation with respect to the "bulge"
 of the distribution. Thus the angular dependence here comes due to the fact that the {\bf average} density is impact parameter dependent and it is the differences in the density that are probed by the dipole.

It is of course not necessary that the function $P(\rho)$ has one dominating maximum. One can imagine several broad maxima, corresponding to several centers of density distribution.
On such a target again the angular dependence appears through the dipoles whose size is of the order of the distance scale on which the density varies. The amplitudes of small dipoles are not correlated in this approximation. In other words, this mechanism of the angular correlations looks very different from the one we have discussed above. The difference is not just qualitative. The length scale that is relevant for the mechanism of Ref.(\refcite{LR}) is not the correlation length of the color electric fields, but rather the correlation length of the density. In other words the correlations should be maximal not at $k\sim Q_s$, but rather at $k\sim l^{-1}_{Q_s}$, where $l_{Q_s}$ is the correlation length of the saturation momentum in the transverse plane. Although naively one might think that the two are of the same order, this does not have to be case. Indeed in a model considered in Ref.(\refcite{muellermunier}) the latter correlation length is parametrically  larger, in the high energy limit  $l_{Q_s}\sim Q^{-1}_se^{c\ln^2(1/\alpha_s)}$. This mechanism would then lead to correlations at very small transverse momentum.

\section{Conclusions}

We have explained that rapidity and angular correlations are a very general feature of particle production at high energy. They are an almost automatic consequence 
of the boost invariance of the projectile wave function, provided  the particle density/scattering probability is large enough so that there is non-negligible probability to produce more than one particle at fixed impact parameter within the saturation radius $1/Q_s$. 

There are good reasons to expect that the factorization of both projectile and target averages is broken {\it at leading order in $1/N_c$} in the kinematical domain relevant to the correlated production of particles. To study this question one certainly has to go beyond simple rotationally invariant solutions of the BK equation, but more importantly, include Pomeron loops in high energy evolution. While technically challenging, it would be very interesting to understand and quantify this effect. 
 
We expect the correlated production to be maximal at momenta close to the saturation momentum $Q_s$. It probes the hadronic wave function at transverse resolutions below $1/Q_s$, and should be dominated by $r\sim 1/Q_s$ for trivial phase space reasons. To be able to describe fluctuations at high energy quantitatively one needs to understand the evolution of the high momentum modes of the target fields in the dilute regime and the semihard momenta in the regime of the onset of saturation. In a sense one has to include the KLWMIJ evolution for the high momentum modes and JIMWLK evolution for the low momentum ones within the same framework. The situation is even more complicated, since one needs to understand the structure of the wave function around the saturation momentum, where the Pomeron merging and splitting (JIMWLK and KLWMIJ contributions) are of equal importance and one cannot just roughly switch from one description to the other.  
In other words, the interesting momentum range is sensitive at the leading order to the Pomeron loop contributions. 

Current numerical implementations Ref.(\refcite{ddgjlv,DV,DV1}) do not include effects of KLWMIJ evolution at high momenta, nor Pomeron loops at $Q_s$, and as a result only include $1/N_c$ suppressed contributions to the correlated emission. In our view it is crucial to understand the effects mentioned above and include them in numerical calculations in order to have a theoretically robust and reliable description of correlated particle emission at high energy.

\section*{Acknowledgments}
The work of AK is supported by DOE grant DE-FG02-92ER40716.
The work of ML is partially supported by the Marie Curie Grant  PIRG-GA-2009-256313 and  THE ISRAEL SCIENCE FOUNDATION.
M.L. thanks  UCONN's  Physics Department for the hospitality during which this work was initiated.
\appendix

\section{Derivation of single and double inclusive production probabilities}

In this Appendix, we sketch the derivation of Eqs.~(\ref{single},\ref{double}).
We start from the wave-function (\ref{wf}) and follow Refs. (\refcite{baier,KLincl}) for the derivation.
Two observables of the interest are 
\begin{eqnarray}
\hat{\cal O}_g&=&a^{\dagger\,a}(k,\eta)\,a^a(k,\eta)\, =\,\int d^2z\,d^2\bar z\,e^{ik(z-\bar z)}\,a^{\dagger\,a}(z,\eta)\,a^a(\bar z,\eta)\ \nonumber \\
\hat{\cal O}_{2g}&=&a^{\dagger\,a}(k,\eta)\,a^a(k,\eta)\,a^{\dagger\,b}(p,\xi)\,a^b(p,\xi)=\,
\int _{z\,\bar z}\,e^{ik(z-\bar z)}\,a^{\dagger\,a}(z,\eta)\,a^a(\bar z,\eta)\,\times \nonumber \\
&&\times\,\int_{u\,v}\,e^{ip(u-v)}\,a^{\dagger\,b}(u,\xi)\,a^b(v,\xi)
\end{eqnarray}
At high collision energy, each projectile's gluon scatters off the target with the eikonal $S$-matrix 
\beq
\langle a^{\dagger\,b} (x,\eta)|\hat S|a^a(x,\eta)\rangle\,=\,S^{ab}(x)\,=\Big[{\cal P}\,\exp\{i \,T^c\,\alpha^c_t(x)\}\Big]^{ab}
\eeq
In the light cone gauge ($A^-=0$), $\alpha_t(x)=\int dx^-\,A^+(x)$ is a field of the target on which the gluons scatter. 
To compute one of the observables $\hat{\cal O}$ in the final state, we have to include initial and final state radiation as encoded in 
$\Omega$ (see \ref{wf}), to account for decoherence as a result of collision process, and finally to average over all possible configuration
of projectile and target. The result have the following form
\beq
\langle \hat{\cal O}\rangle_{P,T}\,=\,\langle\psi|\, \Omega^{ \dagger}\,\,(1\,-\,\hat S^\dagger)\,
\Omega\,\,\,\hat{\cal O}\,\,\,
\Omega^{ \dagger}\,\, 
\,(1\,-\,\hat S)\,\Omega\,|\psi
\rangle_{P,T}
\eeq
and after trivial algebra over the soft gluons we arrive for single inclusive production to Eq. (\ref{single}).
For the double inclusive the final result reads
\begin{equation}
\frac{dN}{ d^2pd^2kd\eta d\xi}=\langle A_{ij}^{ab}(k,p)A_{ij}^{*ab}(k,p)\rangle _{P,T}
\label{section}
\end{equation}
with the amplitude
\begin{eqnarray}
&&A_{ij}^{ab}(k,p)=
 \int_{u,z}e^{ikz+ipu}\int_{x_1, x_2} \\
 &&\times
     \Big\{f_i( z -  x_1)
     \left[S( x_1)-S( z)\right]\rho(x_1)\Big\}^a\Big\{f_j(u - x_2)\left[ S( u)- S(x_2)\right]\rho(x_2)\Big\}^b\nonumber\\
&&-\frac{g}{ 2} \int_{x_1} 
f_i(z -  x_1)
  f_j(u-  x_1)
  \Big\{\left[S( x_1)-S( z)\right]\tilde\rho( x_1)
  \left[ S^\dagger(u)+S^\dagger(x_1)\right]\Big\}^{ab}\nonumber\\
&& + g\int_{ x_1} 
f_i( z - u)
 f_j( u - x_1)
     \left\{ \left( S( z) - S( u) \right)
        \tilde\rho( x_1) S^{\dagger}( u)\right\}^{ab} \, \nonumber .\label{amplitude}
\end{eqnarray}
and we have defined $\tilde\rho\equiv -i T^a\rho^a$. 
%The charge density is normalized such that for a single gluon $\rho^a=gT^a$.
%In these formulae $\rho^a(x)$ is the valence color charge density in the projectile wave function, while $S^{ab}(x)$ is the eikonal scattering matrix 
%determined by the target color fields.
%The average in eq.(\ref{section}) denotes averaging over the projectile and the target wave functions. 
We  note that in this expression the gluon with momentum $p$ is assumed to have larger rapidity, and thus the emission of the two gluons is not completely symmetric.

In Eq. (\ref{double})  we keep only the term with four $\rho$.
This term is leading in the limit of large color  density $\rho\sim 1/g$. One should keep in mind, however that in this limit other terms not included in 
Eq.~(\ref{amplitude}) are equally important\cite{raju1,raju2,KLploop,KLploop1}. The second term in Eq.~(\ref{amplitude})
corresponds to production of two gluons emitted from the same color source in the incoming projectile wave-function. The third term corresponds to the process whereby the softer gluon has been emitted in the wave function by the harder one, with both gluons subsequently produced in the collision. In terms of BFKL ladders, the (square of the) first term is a part of the diagram containing two independent ladders, while the (square of the) last two terms describe emission of two gluons contained in the same BFKL ladder.

%\begin{thebibliography}{000} %for 3 digits
%\begin{thebibliography}{00}  %for 2 digits

\end{document}